\documentclass[%
 reprint,
superscriptaddress,
 amsmath,amssymb,
 aps,
]{revtex4-1}

\usepackage{graphicx}% Include figure files
\usepackage{dcolumn}% Align table columns on decimal point
\usepackage{bm}% bold math
\usepackage{comment}
\usepackage{textcomp}
\begin{document}

%\preprint{APS/123-QED}

\title{Optical coherence of Er$^{3+}$:Y$_2$O$_3$ ceramics for telecommunication quantum technologies}% Force line breaks with \\
\author{Rikuto Fukumori}
\email{These authors contributed equally to the experiments.}
\affiliation{Pritzker School of Molecular Engineering, University of Chicago, Chicago, Illinois, 60637, USA}
\author{Yizhong Huang}
\email{These authors contributed equally to the experiments.}
\affiliation{Pritzker School of Molecular Engineering, University of Chicago, Chicago, Illinois, 60637, USA}
\author{Jun Yang}
\affiliation{Corning Research \& Development Corporation, Sullivan Park, Painted Post, NY 14870, USA}
\author{Haitao Zhang}
\affiliation{Corning Research \& Development Corporation, Sullivan Park, Painted Post, NY 14870, USA}
\author{Tian Zhong}
\email{Corresponding author: tzh@uchicago.edu}
\affiliation{Pritzker School of Molecular Engineering, University of Chicago, Chicago, Illinois, 60637, USA}

\date{\today}% It is always \today, today,
             %  but any date may be explicitly specified

\begin{abstract}
We report an optical homogeneous linewidth of 580 $\pm$ 20 Hz of Er$^{3+}$:Y$_2$O$_3$ ceramics at millikelvin temperatures, narrowest so far in rare-earth doped ceramics. We show slow spectral diffusion of $\sim$2 kHz over a millisecond time scale. Temperature, field dependence of optical coherence and spectral diffusions reveal the remaining dephasing mechanism as elastic two-level systems in polycrystalline grain boundaries and superhyperfine interactions of Er$^{3+}$ with nuclear spins. In addition, we perform spectral holeburning and measure up to 5 s hole lifetimes. These spectroscopic results put Er$^{3+}$:Y$_2$O$_3$ ceramics as a promising candidate for telecommunication quantum memories and light-matter interfaces.
\end{abstract}

\maketitle

%\tableofcontents

\section{\label{sec:level1}Introduction}
Quantum light-matter interfaces connecting optical and spin degrees of freedom are important building blocks for quantum interconnect technologies. Their potential applications include optical quantum memories \cite{Lauritzen2011} for long-distance quantum networks \cite{kimble2008}, optical-microwave quantum transductions \cite{Kurizki2015}, and nanoscopic quantum sensing \cite{Degen2017}. A strong material candidate of quantum light-matter interfaces are rare-earth ion doped solids. The shielded 4f-4f intrashell transitions of rare-earth dopants give rise to exceptionally long optical and spin coherence lifetimes \cite{Thiel2011Lumin, Equall1994, MZhong2015, Bottger2009, Rancic2017}.

In particular, erbium (Er$^{3+}$) is a paramagnetic ion with a half-integer spin (Kramers ion) with an optical transition in the low-loss telecommunication C-band around 1530-1540 nm. When doped into host matrices with low magnetic noise, exceptionally narrow optical homogeneous linewidth ($\Gamma_h$) of 73 Hz \cite{Bottger2009} and a long hyperfine spin coherence time of 1.3 s \cite{Rancic2017} have been measured in $\text{Er}^{3+}$:$\text{Y}_2\text{Si}\text{O}_5$ and ${}^{167}\text{Er}^{3+}$:$\text{Y}_2\text{Si}\text{O}_5$, respectively. The telecom-fiber compatibility paired with long-lived coherence makes Er$^{3+}$ an appealing choice for realizing quantum memories in a fiber-based quantum network, as it eliminates the need for photonic frequency conversion. Furthermore, Er$^{3+}$ has a large gyromagnetic ratio, up to g=15 in $\text{Er}^{3+}$:$\text{Y}_2\text{Si}\text{O}_5$ \cite{Sun2008} and g=12 in $\text{Er}^{3+}$:$\text{Y}_2\text{O}_3$ \cite{Harris2001}, which allow efficient coupling to microwave fields for microwave-optical transduction or coupling to other magnetic targets for sensing applications. On the other hand, the strong magnetism of Er$^{3+}$ imposes significant challenges to attain long coherence lifetimes due to undesired interactions with other Er$^{3+}$ ions, phonons, and impurities in the host. Consequently, previous results of long coherence lifetimes were mostly obtained by applying a very large field of a few teslas to strongly suppress these interactions \cite{Bottger2009, Rancic2017}. Alternatively, adverse Er$^{3+}$-Er$^{3+}$ spin flip-flops and Er$^{3+}$-phonon interactions can be suppressed by freezing Er$^{3+}$ electronic spins at dilution temperatures. This latter technique has been applied to $\text{Er}^{3+}$:$\text{Li}\text{Y}\text{F}_4$ \cite{Kukharchyk2018} and $\text{Er}^{3+}$:$\text{Y}_2\text{Si}\text{O}_5$ single crystals \cite{Craiciu2019}.

Compared to $\text{Er}^{3+}$:$\text{Y}_2\text{Si}\text{O}_5$ which has been well studied in the literature \cite{Bottger2006, Bottger2009}, $\text{Y}_2\text{O}_3$ is another promising host material for Er$^{3+}$ with low nuclear magnetic moments. One advantage of $\text{Y}_2\text{O}_3$ is its simpler chemical composition and lattice structure, which allows synthesis of micro/nano-structured $\text{Y}_2\text{O}_3$ in different topologies using bottom-up approaches \cite{TZhong2019}, such as thin films \cite{Scarafagio2019}, nanoparticles \cite{Bartholomew2017}, and ceramics \cite{Zhang2017}, with increasingly good control of the material volume, particle sizes, and doping concentrations. Furthermore, rare-earth doped $\text{Y}_2\text{O}_3$ has already shown promising coherence characteristics. For instance, near radiatively-limited optical homogeneous linewidth of  $\Gamma_h$=760 Hz was reported in single crystal $\text{Eu}$:$\text{Y}_2\text{O}_3$ \cite{Macfarlane1981}. Additionally, $\Gamma_h$=85.6 kHz has been reported for Eu:Y$_2$O$_3$ nanoparticles \cite{Perrot2013} and $\Gamma_h$=4 kHz for transparent ceramics \cite{Kunkel2016}. Recently, a narrow inhomogeneous linewidth of 430 MHz and $\Gamma_h$=11.2 kHz in the telecom band have also been demonstrated in $\text{Er}^{3+}$:$\text{Y}_2\text{O}_3$ \cite{Zhang2017}. To ascertain the full potential of $\text{Er}^{3+}$:$\text{Y}_2\text{O}_3$ for quantum technologies, an in-depth coherent spectroscopy at dilution temperatures is thus desirable. An understanding of dephasing mechanisms in $\text{Er}^{3+}$:$\text{Y}_2\text{O}_3$ will also provide insights for optimization of the synthesis processes to realize materials with tailored quantum characteristics.

In this work, we measure optical coherence properties of the ${}^4I_{13/2}(Y_1)\rightarrow {}^4I_{15/2}(Z_1)$ telecom transition in transparent $\text{Er}^{3+}$:$\text{Y}_2\text{O}_3$ ceramics in the $<$100 mK temperature regime and obtain the narrowest optical homogeneous linewidth measured so far in rare-earth doped ceramics of 580 Hz at an applied magnetic field of 0.7 T. The measured transition shows a slow spectral diffusion of $\sim$2 kHz over 1 ms. We systematically investigate field, temperature dependence of $\Gamma_h$ and spectral diffusions to determine the dephasing mechanisms, and quantify the remaining broadening due to coupling to elastic two-level systems (TLS) and nuclear spins. We also measure spectral hole lifetimes up to 5 s. Our results of sub-kilohertz optical linewidth with a slow spectral diffusion of $\text{Er}^{3+}$:$\text{Y}_2\text{O}_3$ demonstrates a significant potential of this material for telecommunication quantum technologies.

\section{\label{sec:level1}Experimental Methods}
The material under study is transparent 20 parts per million (ppm) doped $\text{Er}^{3+}$:$\text{Y}_2\text{O}_3$ ceramics, with dimensions of 9.5 mm x 3 mm x 1.7 mm. This material is made by sintering Er$^{3+}$ doped Y$_2$O$_3$ nanoparticles (40 nm) in the following way. The nanoparticles are pressed into a pellet in a steel die at approximately 8 Klbs force. This pellet is isostatically pressed at 25 Kpsi in a latex iso-pressing sheath at room temperature, followed by sintering at 1500 \textdegree{}C in air for 2 hours. The pellet is then hot isostatically pressed at 1490 \textdegree{}C for 16 hours at 29 Kpsi under an argon atmosphere in a graphite furnace. The pellet is buried inside Y$_2$O$_3$ powder during HIP to reduce carbon contamination which comes from the graphite furnace.  The surface of the sample is polished with a roughness of about 5 nm. This resulted in polycrystals with an average cross-sectional grain size of 0.3 $\mu$m$^2$. Further details of the sample and manufacturing process are outlined in \cite{Zhang2017}.

A fiber-coupled telecom diode laser (Toptica CTL 1500) was used for optical measurements. Two cascaded fiber-based acousto-optic modulators (AOM) at 200 MHz each were used to modulate the frequency and intensity of the laser, giving a 100 dB extinction ratio. The light is sent through a fiber-based circulator, which separated the input and output light from the sample.

The sample was mounted on the mixing chamber stage (MXC) of a dilution refrigerator, and a 3-axis nanopositioner was used for sample optical alignment. Two copper clips were used to secure the sample on a copper stage for increased thermal conductivity. The light exiting the fiber was focused onto the sample with an aspheric doublet, down to a spot size of 12 $\mu$m, and reflected off the gold-coated sample stage and back into the fiber. This created a double pass configuration through the sample. The percentage of the light exiting the fiber to the light collected back by the fiber was about 30 \%. A superconducting 6-1-1 T vector magnet was mounted around the sample. Because the sample is ceramics and has no preferred crystallographic directions, the magnet was operated only along one (6 T) axis.

For lifetime measurements, photoluminescence (PL) was detected by sending a 1.5 $\mu$s excitation pulse, gated by two AOMs, and the reflection detected by a superconducting nanowire single photon detector (SNSPD). The dark count rate of the SNSPD was measured to $50\pm 10$ Hz, and the detector efficiency was about 80\%. For absorption spectroscopy, the light was sent to the sample without AOMs, and the reflection was measured with a photodiode.

For optical coherence spectroscopy, two pulse and three pulse photon echoes with heterodyne detection was used. The light before the AOM was split with a fiber-based 50:50 beamsplitter, and one path was sent through an AOM and to the sample, while the other path was used as the local oscillator (LO). The reflection from the sample and the LO was recombined with another 50:50 fiber-based beamsplitter, and the beating at 200 MHz was observed with a photodiode. A variable attenuator and polarization controller was used for the LO path to optimize the echo intensity.

\section{\label{sec:level1}Results}

\begin{figure*}[t]
  \centering
    \includegraphics[width=0.75\linewidth]{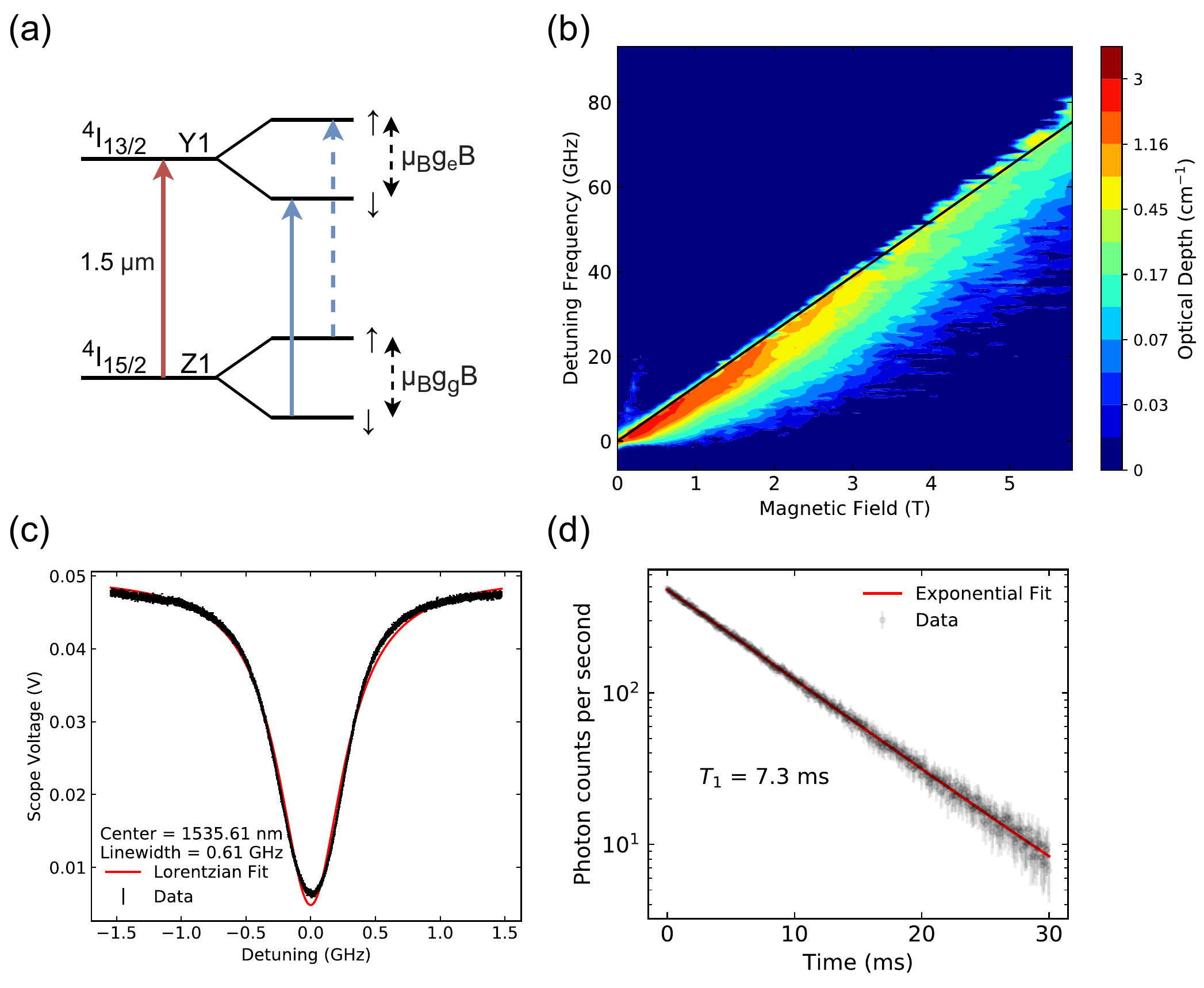}
  \caption{(a) Energy level diagram for the crystal field ${}^4I_{13/2}(Y_1)\rightarrow {}^4I_{15/2}(Z_1$) transition (red arrow) for Er$^{3+}$. An external magnetic field splits both levels into Zeeman doublets. At low temperatures, most ions occupy the spin down levels, thus the dominant transition is indicated by the solid blue arrow. (b) Emission spectra as a function of magnetic field and frequency. The lineshape gradually broadens with increasing field. The slope of a contour is proportional to $(g_g-g_e)_z$ and the solid black line indicates the line of maximum absorption, with $(g_g-g_e)_z=1.84$. (c) Zero-field inhomogeneous linewidth measured by absorption spectroscopy. The data are fitted by a Lorenztian, with a full-width-at-half-maximum of 0.61 GHz. (d) Optical lifetime, measured with PL detected by a SNSPD, integrated over 10 seconds. The curve is fitted to a single exponential decay, giving a $T_1=7.3$ ms.}
\end{figure*}

\subsection{\label{sec:level2}Optical absorption spectroscopy}
Figure 1(a) shows the energy diagram for the ${}^4I_{13/2}(Y_1)\rightarrow {}^4I_{15/2}(Z_1)$ optical transition in the $C_2$ crystal symmetry site. Application of a magnetic field splits each of these crystal field levels into two Zeeman levels. The g-tensor for Er$^{3+}$:Y$_2$O$_3$ crystals was measured in \cite{Harris2001}. For polycrystalline ceramics, the Er$^{3+}$ ions are randomly oriented, and thus experience different g-factors. We see this in the broadening of the inhomogeneous linewidth with applied magnetic field, as shown in Fig.~1(b). Groups of ions with different g-factors spread out in the frequency domain, due to a distribution of $|g_g-g_e|$ values. For the remainder of this paper, we measure optical coherence and spin lifetimes at frequencies corresponding to maximum optical absorption, as indicated by the black line in Fig.~1(b).

The zero-field absorption spectrum for the Y$_1$-Z$_1$ transition is shown in Fig.~1(c). This yielded a center frequency of 195227.0 GHz and a peak absorption of 87\% for an effective crystal length of 3.4mm, corresponding to an absorption coefficient of 3.0 cm\textsuperscript{-1}. The zero field optical inhomogeneous linewidth was 610 MHz full-width-at-half-maximum, which is an order of magnitude narrower than those reported in similar europium doped ceramics \cite{Ferrier2013}, and still narrower than previously reported $\text{Er}^{3+}$:$\text{Y}_2\text{O}_3$ single crystals \cite{Thiel2011Lumin}. This indicates low disorder and high purity of the ceramics.

The oscillator strength $f$ is given by \cite{Bartolo1968, Henderson2006}

\begin{equation}
    f=4\pi\epsilon_0 \frac{9m_ecn}{\pi e^2 N (n+2)^2}\int\alpha(\nu)d\nu
\end{equation}

\noindent where $\epsilon_0$ is the vacuum permittivity, $e$ is the electron charge, $m_e$ is the electron mass, $c$ is the speed of light, $n$ is the index of refraction, $N$ is the number density of the dopant, and $\alpha$ is the absorption coefficient. Given $\alpha(\nu)$ measured in Fig.~1(c) and the index for Y$_2$O$_3$ at 1535 nm n=1.9, we calculate $f=2.9*10^{-7}$. From this, we calculate the spontaneous emission time $T_{spon}$ for the Y$_1$-Z$_1$ transition using \cite{Liu2010}:

\begin{equation}
    T_{spon}=\frac{m_e\epsilon_0c^3}{2\pi n^2e^2\nu^2f}
\end{equation}
which gives $T_{spon}=34.2$ ms.

Figure 1(d) shows a PL decay integrated over ten seconds. The plot is fitted to a single exponential, yielding an excited state lifetime ($T_1$) of 7.3 ms. We confirmed that reabsorption of emitted photons due to large optical depths was not causing lengthening of the lifetime, by measuring the lifetimes at detuned frequencies where the optical depth was lower. From the measured optical lifetime, we obtain a branching ratio for the Y$_1$-Z$_1$ transition as $T_1/T_{spon}=0.215$.

\subsection{\label{sec:level2}Optical coherence}
We measure the optical coherence using two-pulse photon echoes width heterodyne detection. The $\pi$ pulse was 500 ns long. The inset of Fig.~2 shows typical photon echo decays at various applied fields, along the black line in Fig.~1(b). At all fields, we observed non-exponential decays, which we fit with the Mims decay \cite{Mims1968}
\begin{equation}
    E(\tau)=E_0e^{-(2\tau/T_M)^x}
\end{equation}
\noindent where $\tau$ is the delay between the two pulses, $x$ is the parameter describing spectral diffusion, and $T_M$ is the phase memory time. From the fit, we extract the effective homogeneous linewidth as $\Gamma_{h,eff}=1/(\pi T_M)$. The narrowest linewidth of 580 Hz was observed at 0.7 T.

\begin{figure}
  \centering
    \includegraphics[width=90mm]{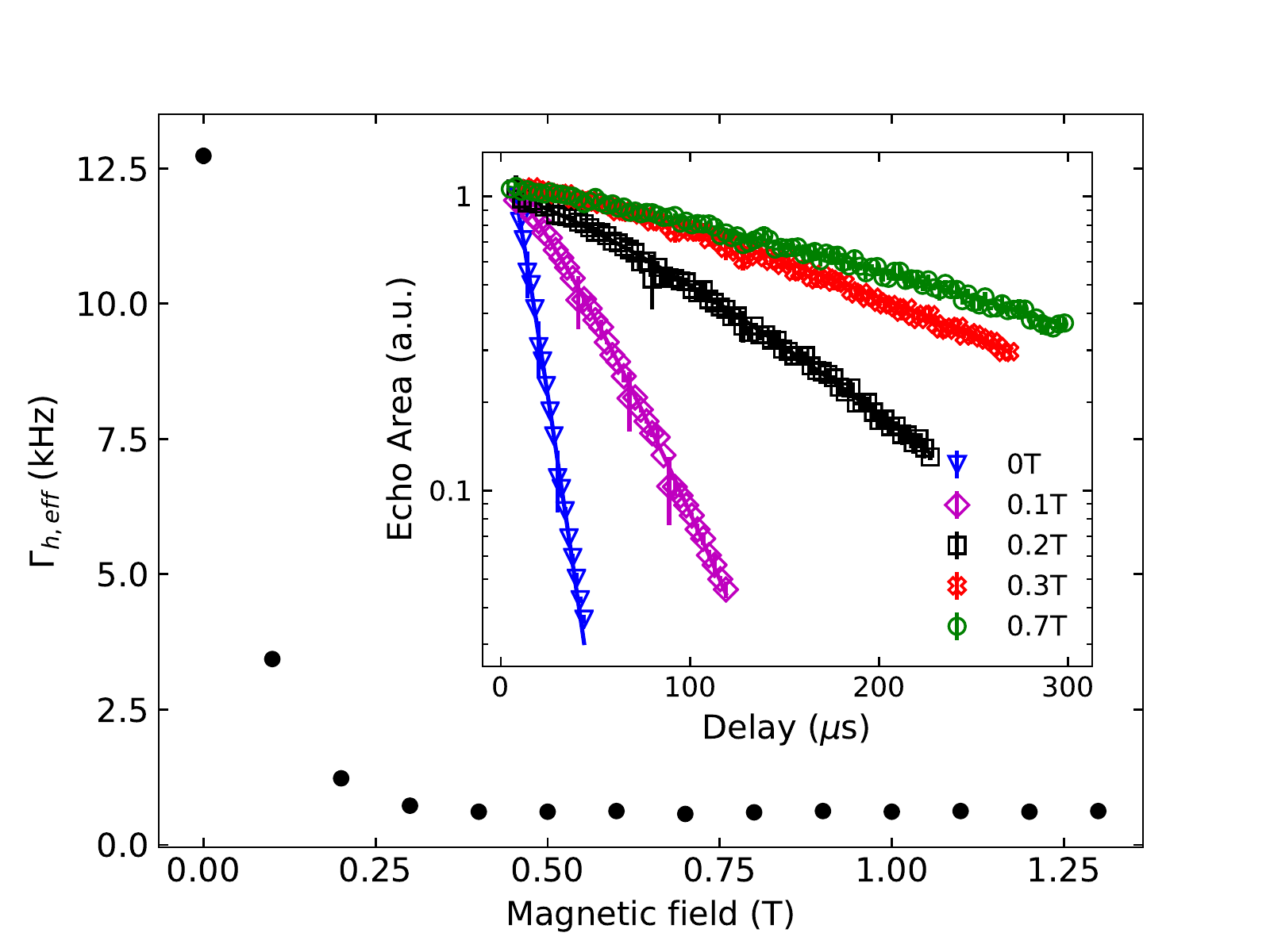}
  \caption{Magnet field dependence of the effective homogeneous linewidth. We see the linewidth decrease from 12.5 kHz at 0 T down to 580 Hz beyond 0.7 T, where the linewidth saturates due to decoherence caused by TLS and superhyperfine interactions. Inset: Normalized echo decays at various fields.}
\end{figure}

To determine the dephasing mechanisms, we break down the contributing factors to $\Gamma_h$ as:

\begin{equation}
    \Gamma_h=\Gamma_{pop} + \Gamma_{ion-ion} + \Gamma_{ion-spin} + \Gamma_{TLS} + \Gamma_{phonon}
\end{equation}

$\Gamma_{pop}$ is the contribution from the excited state radiative lifetime. This gives $\Gamma_{pop}=1/(2\pi * T_1)=21.8$ Hz, a negligible contribution to the overall linewidth. $\Gamma_{ion-ion}$ includes contributions from two factors. One is instantaneous spectral diffusion (ISD) due to strong optical excitations that can abruptly change the local environment \cite{Liu1990}. This would cause dephasing with increasing power of the second pulse $\pi$. However, we saw no power dependence of the linewidth, suggesting that ISD does not contribute. Furthermore, previous work on a similar sample has shown no change in the linewidth between 2.0 ppm and 11.5 ppm doped samples, further indicating that ISD does not contribute \cite{Zhang2017}. The second contribution to $\Gamma_{ion-ion}$ is the resonant Er$^{3+}$-Er$^{3+}$ flip-flops \cite{Car2018}. From Fig. 1(b), we saw minimal population in the upper Zeeman state at B$>$0.1 T. Due to thermal depopulation of the upper Zeeman state, we expect the dephasing from Er$^{3+}$-Er$^{3+}$ flip-flops to be small.

$\Gamma_{ion-spin}$ includes dephasing due to the superhyperfine interaction of Er$^{3+}$ with Y$^{3+}$ nuclear spins, and is often the limiting dephasing mechanism, known as the superhyperfine limit \cite{Kukharchyk2018}. This contribution is usually small, owing to the low nuclear magnetic moment of Y$^{3+}$, and the large magnetic moment of Er$^{3+}$ creating a frozen core of yttrium whose flipping rates are significantly slowed \cite{Kukharchyk2018}. Since the magnetic noise of Y$^{3+}$ is independent of field, the characteristic of the superhyperfine limit is a saturation of the homogeneous linewidth beyond a certain field. $\Gamma_{ion-spin}$ also includes contributions from hyperfine interactions between the $^{167}$Er$^{3+}$ (22.8\% natural abundance) electron spin and its nuclear spin of 7/2 . Additionally for ceramics, there may be magnetic impurities added during the manufacturing process that could increase $\Gamma_{ion-spin}$.

$\Gamma_{TLS}$ is the dephasing that arises from fluctuations in the local environment (e.g.~magnetic, electric fields) due to tunneling between two configurations with similar energy, known as tunneling two-level systems (TLS) \cite{Anderson1972, Phillips1972}. For Er$^{3+}$-doped systems with TLS, two types of TLS have been observed \cite{Macfarlane2007}. One is the elastic TLS modes, which are independent of magnetic field. Another is the coupling between the large anisotropic magnetic moment of Er$^{3+}$ and the elastic TLS modes, facilitated through elastic-dipole interactions \cite{Huber1984}, which we refer to as magnetic TLS. Magnetic TLS noise is expected to decrease with applied magnetic field, as opposed to elastic TLS which is field independent \cite{Macfarlane2007}. $\Gamma_{TLS}$ is often the dominating contributor in amorphous solids \cite{Kunkel2016, Black1977, Breinl1984}, although to a lesser degree in ceramics as compared to glasses. Effects of TLS has been observed in other ceramics \cite{Ferrier2013}, and is likely contributing to this material as well. $\Gamma_{phonon}$ includes spin relaxation caused by three primary phonon scattering processes: the direct one-phonon process, and the two-phonon Raman and Orbach processes \cite{Bottger2006, Sun2012}.

\subsection{\label{sec:level2}Field dependence of $\Gamma_h$}

$\Gamma_h$ at increasing magnetic fields along the maximum absorption line are plotted in Fig.~2. We observed a decrease in linewidth from 12.5 kHz at zero field, to a minimum of 580 $\pm$ 20 Hz at 0.7 T. The uncertainty is extracted from the fit as one standard deviation. The linewidth does not change at higher fields. The reduction of $\Gamma_h$ with field is indicative of decrease in both Er$^{3+}$-Er$^{3+}$ flip flops and magnetic TLS. The subsequent saturation of $\Gamma_h$ is indicative of both the superhyperfine limit and elastic TLS. The measured homogeneous linewidth of 580 Hz is about an order of magnitude narrower than previously measured transparent ceramics \cite{Zhang2017, Kunkel2016}, though it is still broader than the $T_1$-limited value of 22 Hz.

Between magnetic fields of 0.01 to 0.1 T, we observed periodic modulations in the echo amplitudes with $\tau$. These oscillations decreased in amplitude and increased in frequency as the field was increased from 0.01 to 0.1 T. Two pulse echoes done on Er$^{3+}$Y$_2$SiO$_5$ show similar oscillations below 0.1 T \cite{Car2018}. These strong oscillations prevented accurate fitting and extraction of $x$ parameter and $T_M$.

\subsection{\label{sec:level2}Temperature dependence of $\Gamma_h$}
To further investigate the dephasing mechanisms, we measured the temperature dependence of homogeneous linewidth at 0.1 T and 0.7 T, as shown in Fig.~3(a) and (b), respectively. At temperatures under 1 K, we see a linear increase of linewidth with temperature, characteristic of TLS being the dominating dephasing mechanism \cite{Anderson1972, Phillips1972, Flinn1994, Macfarlane2004}, without any increasing contributions from phonons. At higher temperatures ($>$4K), non-linear behavior attributed to phonon scattering processes would be expected \cite{Zhang2017}, but we did not measure in that temperature range due to limitations of our setup. Note that below 100 mK, we see a slight non-linear plateau. This is likely due to the saturation of linewidth caused by coherently coupled TLS pairs \cite{Ding2018}, which dominate once the temperature falls under 100 mK where phonon scattering is very weak. It is also possible that the thermal conduction between the sample and the MXC is reduced at these temperatures, so the actual temperature of the sample deviated from the MXC temperature.

\begin{figure}
  \centering
    \includegraphics[width=0.9\linewidth]{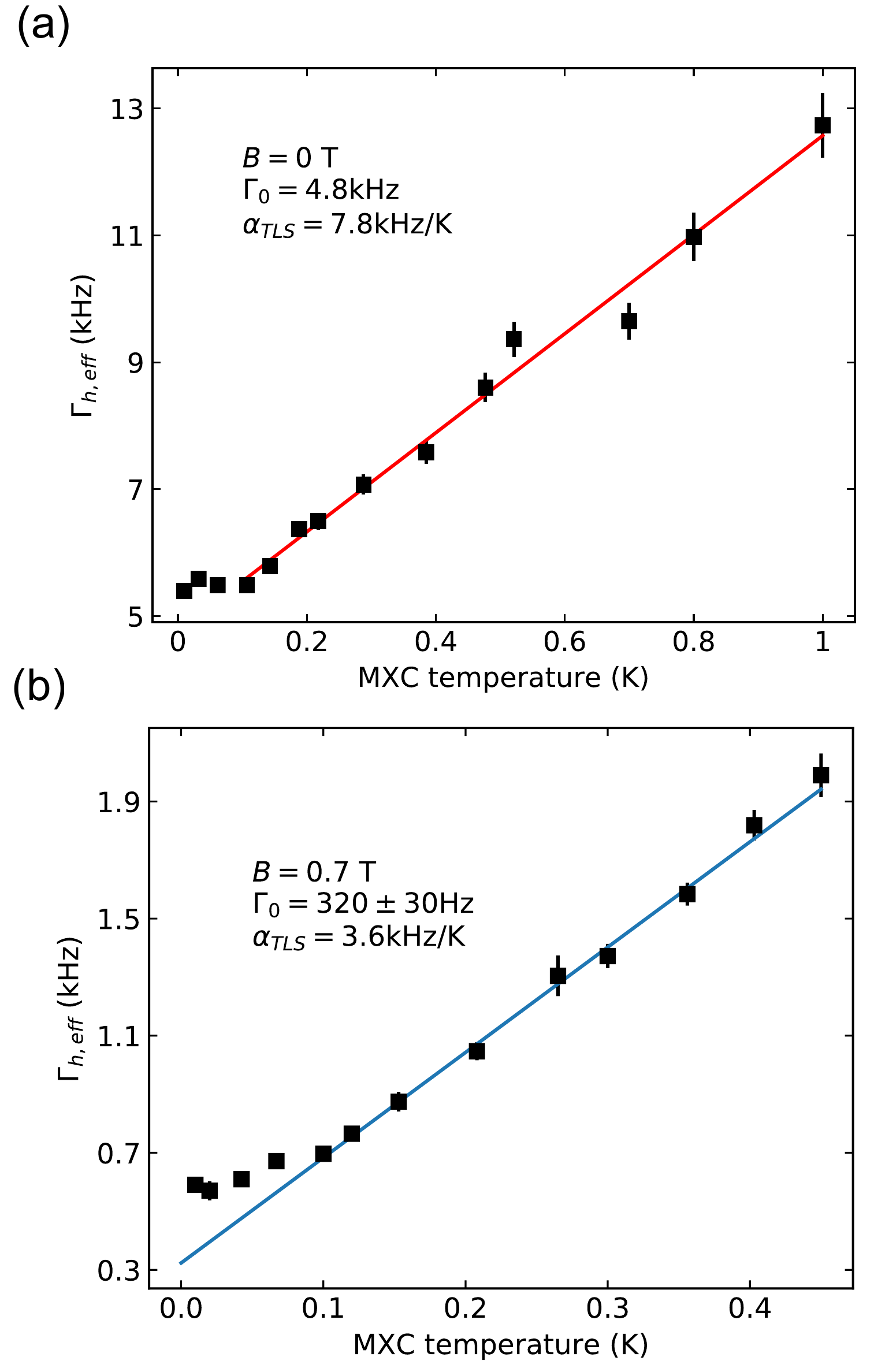}
  \caption{(a) Temperature dependence of the homogeneous linewidth at 0.1 T. The linearity indicates decoherence dominated by TLS. The saturation at temperatures below 100 mK is also characteristic of coupled TLS pairs. (b) The same measurements at 0.7 T. We see decrease in both $\alpha_{TLS}$ and $\Gamma_0$.}
\end{figure}

We perform a linear fit in Fig.~3 (excluding the points below 100 mK), according to:
\begin{equation}
    \Gamma_{eff}=\Gamma_0+\alpha_{TLS}T
\end{equation}
where $\Gamma_0$ is the linewidth at 0 K. We obtain fit parameters of $\Gamma_0=4.8$ kHz and $\alpha_{TLS}=7.8$ kHz/K at 0.1 T, and $\Gamma_0=320$ Hz and $\alpha_{TLS}=3.6$ kHz/K at 0.7 T. We first note that the TLS coupling $\alpha_{TLS}$ is half at the higher field. The reduction in $\alpha_{TLS}$ is likely due to a reduction of magnetic TLS, which have been seen in similar systems \cite{Ding2018, Staudt2006}. Some TLS couples to the strong dipole moment of Er$^{3+}$, and the subsequent application of magnetic field decouples the two energy levels that constitute the TLS. This leads to lower tunneling rate, thus reducing the magnetic noise caused by TLS. We also note that $\Gamma_0=320$ Hz is thus the linewidth in the absence of TLS.

\subsection{\label{sec:level2}Spectral Diffusion}
\begin{figure}
  \centering
    \includegraphics[width=\linewidth]{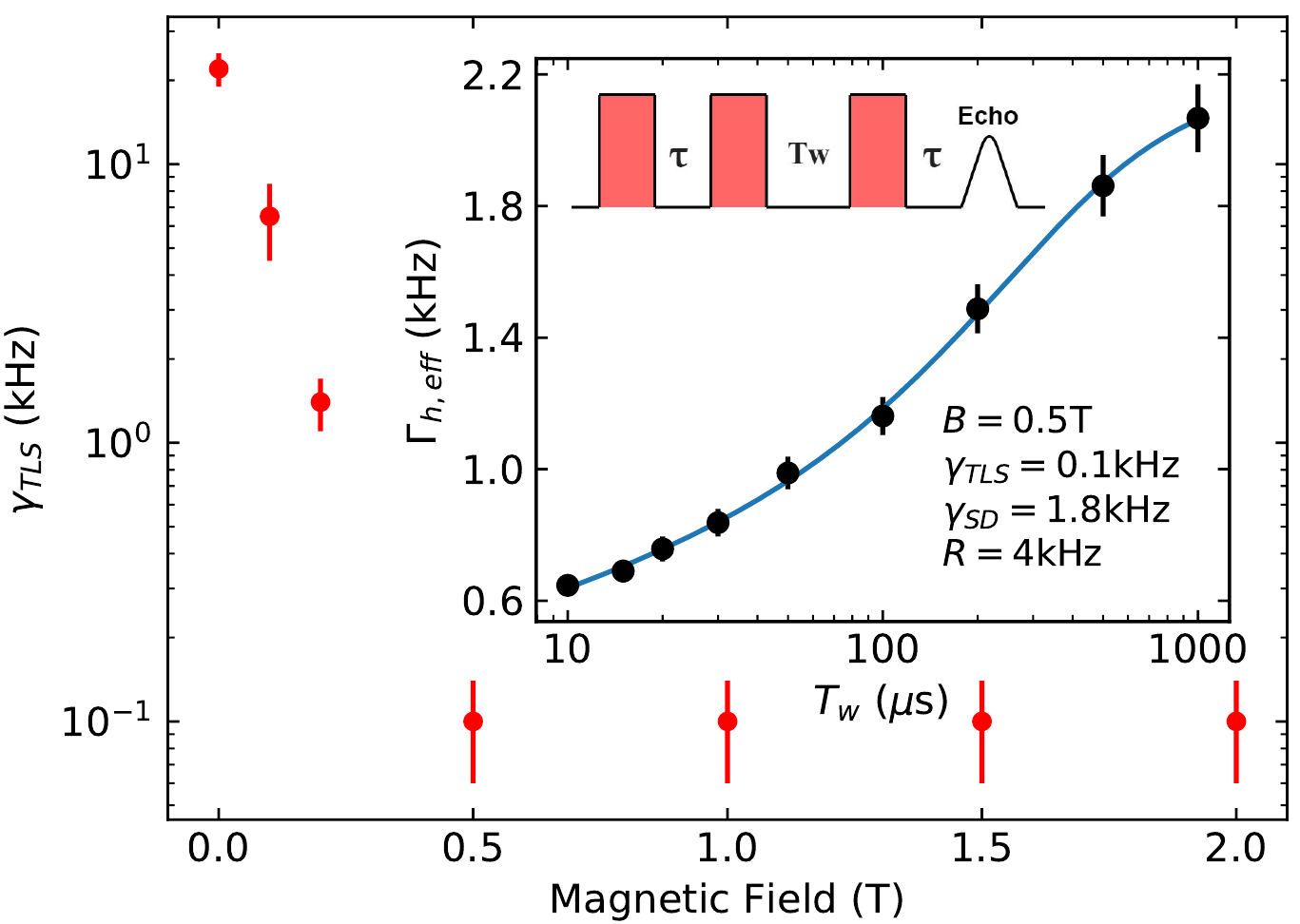}
  \caption{Magnetic field dependence of the $\gamma_{TLS}$, measured with three pulse photon echoes. Inset: Spectral diffusion at 0.5 T, showing broadening from magnetic spin flips and TLS. The blue solid line is a fit to
  Eq. (6).}
\end{figure}

To determine the temporal dependence of the linewidth, we investigate spectral diffusion using stimulated photon echoes (three pulse echoes). Three identical 300 ns pulses were sent, with a delay between the second and third pulse we denote as the wait time $T_w$. The delay between the first and second pulse $\tau$ was varied to extract the coherence time for a given $T_w$.

Spectral diffusion caused by spin flips and TLS is described by
\begin{equation}
    \Gamma_h(T_w) = \gamma_0 + \frac{1}{2}\gamma_{SD}(1-\text{exp}(-RT_w)) +  \gamma_{TLS}\text{log}(\frac{T_w}{t_0})
\end{equation}
where $\Gamma_0$ is the homogeneous linewidth in the absence of spectral diffusion, $\gamma_{SD}$ is the broadening due to magnetic spin flips, $R$ is the Y spin flip rate, $\gamma_{TLS}$ is the TLS coupling strength, and $t_0$ is the minimum measurement timescale, which is 10$\mu$s. Here we take into account the contribution to spectral diffusion from the superhyperfine/hyperfine spin flips, indicated by the exponential term \cite{Bottger2006}, and from TLS, given by the logarithmic term \cite{Black1977, Breinl1984, Littau1992, Silbey1996, Koedijk1996}.

We use Eq.(6) to fit the three pulse echo decays, and the data and fit at $B=0.5$ T is shown in the inset of Fig.~4. We obtain fit parameters of $\gamma_0=0.6\pm0.1$ kHz, $\gamma_{SD}=1.8\pm0.2$ kHz, $R=4.0\pm0.2$ kHz, and $\gamma_{TLS}=0.11\pm0.01$ kHz. At $B=0$ , the fit parameters were $\gamma_0=22$ kHz, $\gamma_{SD}=2.0$ kHz, $R=4$ kHz, $\gamma_{TLS}=20$ kHz. The fitting for  The reduction in $\gamma_{TLS}$ is likely due to a reduction of magnetic TLS, as evident from the temperature dependence study. We saw no significant change in the diffusion rate with further increase in the field beyond 0.5 T. We also note that our value of $\gamma_{TLS}$ at $B=0.5$ T is about an order of magnitude smaller than those measured in Eu$^{3+}$:Y$_2$O$_3$ ceramics \cite{Kunkel2016}, mostly due to a significantly lower temperature in the current measurement.

Next, we consider the x parameter obtained from the Mims fit of the two pulse photon echo decays, which describes spectral diffusion. For all magnetic fields, x was approximately 1.2-1.3. Previous studies on Er$^{3+}$-doped glasses that have shown decoherence dominated by TLS reveal x parameters equal to or less than 1 \cite{Veissier2016, Lutz2017}. Meanwhile, crystalline Er$^{3+}$:Y$_2$O$_3$ at the superhyperfine limit has shown x=1.4 \cite{Thiel2011Lumin}. Therefore, it is probable that x=1.2 arises from a combination of dephasing due to TLS and Er$^{3+}$-Y$^{3+}$ superhyperfine interactions, and the linewidth beyond 0.7 T is limited by both mechanisms.

\subsection{\label{sec:level2}Spectral Hole Lifetime}
The spectral hole lifetime was measured using hole burning, as a function of applied magnetic field. A hole was burnt using a long burning pulse, shaped by an AOM. After some delay, a frequency sweeping probe pulse was sent to measure the amplitude of the hole. The decay of the hole was fitted to an exponential decay, and the decay constant was extracted to obtain hole lifetimes as shown in Fig.~5.

We see two distinct exponentials, one fast decay for wait times less than 8ms, and a slower decay for longer wait times. We attribute the faster decay to the optical $T_1$. This slower decay is likely due to long-lived hyperfine levels in \textsuperscript{167}Er isotopes and is plotted against magnetic field in Fig.~5. We see a small increase in lifetime below 0.2 T field. This is the regime where the electron spin cross-relaxation has been slowed, and the phonon density is still low. As field increases to intermediate values of 0.2-0.6 T, the phonon density increases, suppressing the lifetime. Above 0.8 T, the lifetime increases and plateaus as phonon density is suppressed and the electron spins are frozen, and hyperfine cross-relaxation dominates. This relation is similar to previous studies on Er$^{3+}$:Y$_2$SiO$_5$ \cite{Rancic2017}. At high fields where the narroest $\Gamma_h$ are measured, the frozen electron spin indicates spin-phonon relaxation has negligible contribution to optical dephasing.

\begin{figure}
  \centering
    \includegraphics[width=\linewidth]{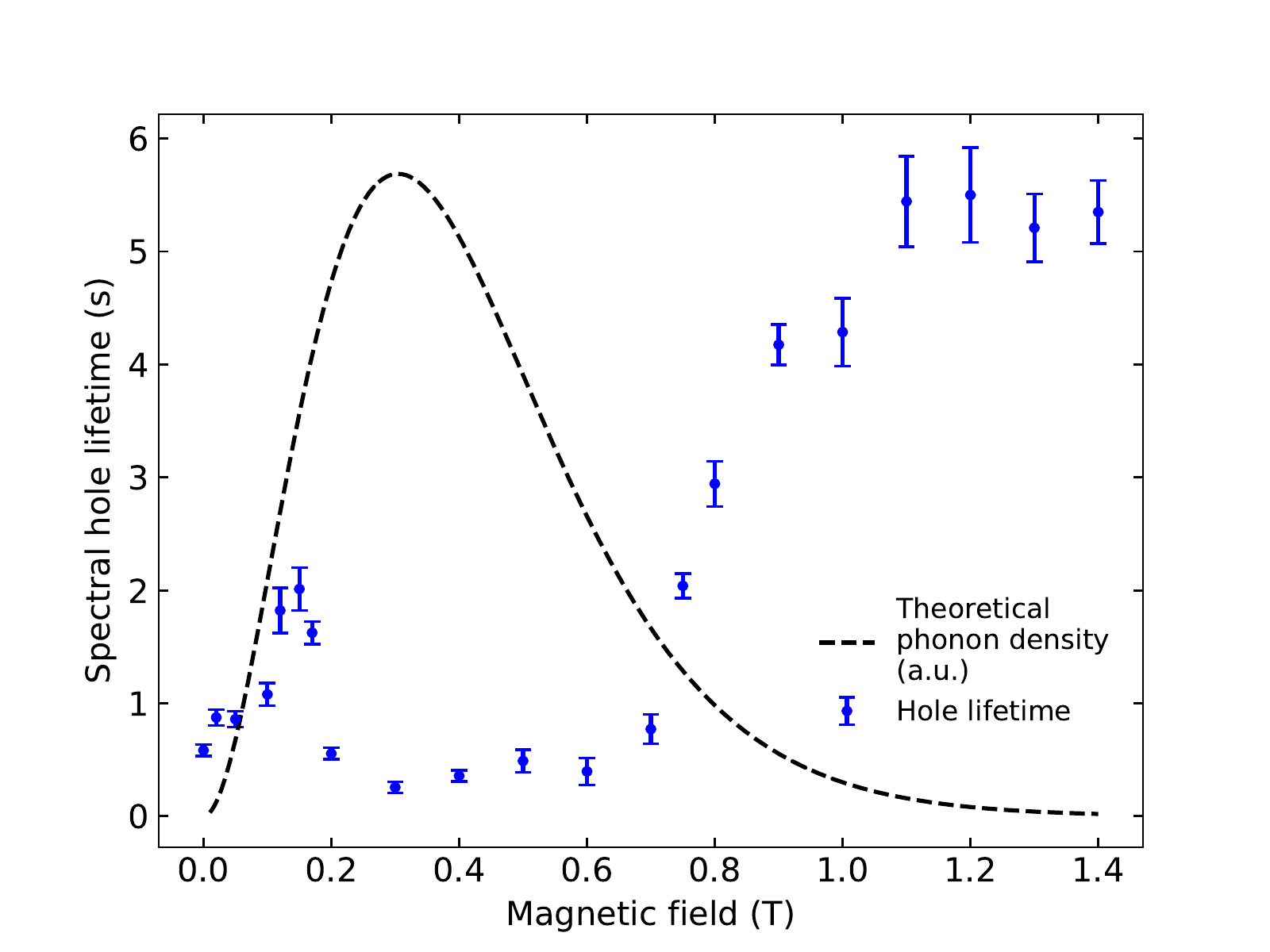}
  \caption{Magnetic field dependence of spectral hole lifetime. The short lifetime between 0.2T to 0.6T indicates increased phonon density. The dotted black line is the black-body phonon density at 0.5 K.}
\end{figure}

A theoretical phonon density curve is overlaid in Fig. 5. The density of states is estimated as black-body radiation with the frequency of the phonons equal to the ground state Zeeman splitting. The mean ground state g-factor $g_g=6.9$ is calculated by taking a spherical average over the 6 equivalent C$_2$ sites, given g-factors measured in \cite{Harris2001}. For these measurements, we estimate that the temperature of the ions is $T=0.5$ K. The increased temperature is likely due to the strong burning pulse, which causes noticeable heating of the sample. We measured the absorption spectrum under similar conditions, and at $B=0.1$ T, we saw about 40 \% of the population in the upper Zeeman state, from which we estimate an actual temperature of 0.5 K. The evident correlation between the phonon density of states and the spectral hole lifetimes suggests that the spin-phonon relaxation plays a role in ceramics similar to that in bulk crystals \cite{Rancic2017}.

\subsection{\label{sec:level2}Summary of dephasing mechanisms}
Here, we provide a summary of the dephasing mechanisms uncovered by the various experiments. At low magnetic fields, there is a combination of dephasing due to TLS, Er$^{3+}$-Er$^{3+}$ interactions, and Er$^{3+}$-Y$^{3+}$ interactions. The application of magnetic field above 0.7 T effectively suppressed the optical dephasing from Er$^{3+}$-Er$^{3+}$ flip-flops, and magnetic TLS.

For ceramics with such a narrow linewidth, it's plausible that charges on the boundaries between crystallite grains contributes some Stark broadening to the homogeneous linewidth. The Stark broadening was significant in Eu$^{3+}$-doped nanoparticles in \cite{Bartholomew2017}, where they estimated a few kHz broadening from the Stark shift from charges on the surface of the nanoparticles. Following the same methods as \cite{Bartholomew2017} to calculate the possible contributions from charges on the grain boundaries given sub-micron sized crystallites, we obtain tens of kHz of broadening. Because of this, we rule out Stark broadening due to charges on crystalline boundaries.

For the remaining linewidth of about 550 Hz above 0.7 T (after subtraction of $T_1$ contribution), from the results of Fig.~3(b), we attribute approximately 230 Hz to elastic TLS. We are left with 320 Hz for Er$^{3+}$-Y$^{3+}$ superhyperfine interactions and possible $^{167}$Er$^{3+}$ hyperfine interactions. Part of the superhyperfine interactions could come from the frozen core, where the large magnetic moment of Er$^{3+}$ slows the flipping of nearby yttrium nuclear spins.

\section{\label{sec:level1}Discussion}
The understanding of relevant dephasing mechanisms inform further works to extend the Er$^{3+}$ coherence lifetimes. With refined synthesis processes, Er$^{3+}$:Y$_2$O$_3$ ceramics with larger grain sizes (e.g. on the order of a few microns) could yield much narrower linewidths as dephasing experienced by ions far from TLSs on the grain boundaries is reduced. An isotopically purified version of this sample with $^{167}$Er$^{3+}$ would allow us to potentially attain long spin coherence lifetimes by performing spin initialization into a non-Boltzmann hyperfine state and suppressing noise due to nuclear spin cross-relaxations \cite{Rancic2017}. The presence of phonon couplings as evident in spectral hole lifetimes in Fig.~5 indicates that phononic band-gap engineering can be used to suppress the phonon density of states in ceramics at spin resonance frequencies \cite{Lutz2016}.

The sub-kilohertz $\Gamma_h$, low spectral diffusion, and long spectral hole lifetime qualifies Er$^{3+}$:Y$_2$O$_3$ ceramics materials for atomic frequency comb (AFC) memories \cite{Afzelius2009}. High storage efficiency and long storage time in the AFC scheme requires that the hole-burned spectral features stay narrow throughout its storage time, which remained difficult to achieve and has been limiting the AFC memories based on $\text{Er}^{3+}$:$\text{Y}_2\text{Si}\text{O}_5$ to date \cite{Lauritzen2011,Craiciu2019}. Similar requirements also apply to controlled reversible inhomogeneous broadening (CRIB), or gradient echo memories \cite{Hedges2010}, for which a sustained narrow optical $\Gamma_h$ and long spectral hole lifetime in Er$^{3+}$:Y$_2$O$_3$ ceramics are advantageous for realizing high performance telecom memories.

We also consider this material for use in single ion quantum devices. Optically addressable single neodymium (Nd$^{3+}$) \cite{Zhong2018} and Er$^{3+}$ \cite{Dibos2018} ions in bulk crystals coupled to nanophotonic cavities have been demonstrated. Main limitations in current systems are significant decoherences for ions in proximity to the crystal surfaces and non-optimal alignment of ions with the photonic cavity fields. The use of sub-micron-sized ceramics crystallites could potentially address these issues, for instance, by embedding a Er$^{3+}$:Y$_2$O$_3$ crystallite in a silicon photonic resonator to achieve deterministic coupling while preserving good optical coherence of individual Er$^{3+}$ ions. Given the already weak optical dephasing, a realistic Purcell enhancement of the Er$^{3+}$ emission as demonstrated in \cite{Dibos2018} will render transform-limited emission from single ions, realizing highly indistinguishable atomic photon sources at the telecom band.

\section{\label{sec:level1}Conclusion}
In this work, we measured spectroscopic properties of polycrystalline $\text{Er}^{3+}$:$\text{Y}_2\text{O}_3$ ceramics for use as quantum light-matter interfaces. We achieved the narrowest optical homogeneous linewidth of 580 Hz of rare-earth doped ceramics, and measured a low spectral diffusion of 2.1 kHz over 1 ms. We further determine that the dominant decoherence mechanisms to be a combination of elastic TLS, superhyperfine interactions between Er$^{3+}$ and Y$^{3+}$ nuclear spins through investigations of field and temperature dependence of the optical homogeneous linewidths as well as spectral diffusions. Our study indicates that transparent $\text{Er}^{3+}$:$\text{Y}_2\text{O}_3$ ceramics is an excellent candidate for developing long-coherence quantum interfaces in the technologically important telecom C-band.

\section*{Acknowledgments}
We are grateful to Charles W. Thiel and John G. Bartholomew for valuable discussions. We acknowledge equipment support from the National Science Foundation EAGER award No. 1843044.

\bibliography{references.bib}

%merlin.mbs apsrev4-1.bst 2010-07-25 4.21a (PWD, AO, DPC) hacked
%Control: key (0)
%Control: author (8) initials jnrlst
%Control: editor formatted (1) identically to author
%Control: production of article title (-1) disabled
%Control: page (0) single
%Control: year (1) truncated
%Control: production of eprint (0) enabled
\begin{thebibliography}{49}%
\makeatletter
\providecommand \@ifxundefined [1]{%
 \@ifx{#1\undefined}
}%
\providecommand \@ifnum [1]{%
 \ifnum #1\expandafter \@firstoftwo
 \else \expandafter \@secondoftwo
 \fi
}%
\providecommand \@ifx [1]{%
 \ifx #1\expandafter \@firstoftwo
 \else \expandafter \@secondoftwo
 \fi
}%
\providecommand \natexlab [1]{#1}%
\providecommand \enquote  [1]{``#1''}%
\providecommand \bibnamefont  [1]{#1}%
\providecommand \bibfnamefont [1]{#1}%
\providecommand \citenamefont [1]{#1}%
\providecommand \href@noop [0]{\@secondoftwo}%
\providecommand \href [0]{\begingroup \@sanitize@url \@href}%
\providecommand \@href[1]{\@@startlink{#1}\@@href}%
\providecommand \@@href[1]{\endgroup#1\@@endlink}%
\providecommand \@sanitize@url [0]{\catcode `\\12\catcode `\$12\catcode
  `\&12\catcode `\#12\catcode `\^12\catcode `\_12\catcode `\%12\relax}%
\providecommand \@@startlink[1]{}%
\providecommand \@@endlink[0]{}%
\providecommand \url  [0]{\begingroup\@sanitize@url \@url }%
\providecommand \@url [1]{\endgroup\@href {#1}{\urlprefix }}%
\providecommand \urlprefix  [0]{URL }%
\providecommand \Eprint [0]{\href }%
\providecommand \doibase [0]{http://dx.doi.org/}%
\providecommand \selectlanguage [0]{\@gobble}%
\providecommand \bibinfo  [0]{\@secondoftwo}%
\providecommand \bibfield  [0]{\@secondoftwo}%
\providecommand \translation [1]{[#1]}%
\providecommand \BibitemOpen [0]{}%
\providecommand \bibitemStop [0]{}%
\providecommand \bibitemNoStop [0]{.\EOS\space}%
\providecommand \EOS [0]{\spacefactor3000\relax}%
\providecommand \BibitemShut  [1]{\csname bibitem#1\endcsname}%
\let\auto@bib@innerbib\@empty
%</preamble>
\bibitem [{\citenamefont {Lauritzen}\ \emph {et~al.}(2011)\citenamefont
  {Lauritzen}, \citenamefont {Min\'a\ifmmode~\check{r}\else \v{r}\fi{}},
  \citenamefont {de~Riedmatten}, \citenamefont {Afzelius},\ and\ \citenamefont
  {Gisin}}]{Lauritzen2011}%
  \BibitemOpen
  \bibfield  {author} {\bibinfo {author} {\bibfnamefont {B.}~\bibnamefont
  {Lauritzen}}, \bibinfo {author} {\bibfnamefont {J.~c.~v.}\ \bibnamefont
  {Min\'a\ifmmode~\check{r}\else \v{r}\fi{}}}, \bibinfo {author} {\bibfnamefont
  {H.}~\bibnamefont {de~Riedmatten}}, \bibinfo {author} {\bibfnamefont
  {M.}~\bibnamefont {Afzelius}}, \ and\ \bibinfo {author} {\bibfnamefont
  {N.}~\bibnamefont {Gisin}},\ }\href {\doibase 10.1103/PhysRevA.83.012318}
  {\bibfield  {journal} {\bibinfo  {journal} {Phys. Rev. A}\ }\textbf {\bibinfo
  {volume} {83}},\ \bibinfo {pages} {012318} (\bibinfo {year}
  {2011})}\BibitemShut {NoStop}%
\bibitem [{\citenamefont {Kimble}(2008)}]{kimble2008}%
  \BibitemOpen
  \bibfield  {author} {\bibinfo {author} {\bibfnamefont {H.~J.}\ \bibnamefont
  {Kimble}},\ }\href@noop {} {\bibfield  {journal} {\bibinfo  {journal}
  {Nature}\ }\textbf {\bibinfo {volume} {453}},\ \bibinfo {pages} {1023}
  (\bibinfo {year} {2008})}\BibitemShut {NoStop}%
\bibitem [{\citenamefont {Kurizki}\ \emph {et~al.}(2015)\citenamefont
  {Kurizki}, \citenamefont {Bertet}, \citenamefont {Kubo}, \citenamefont
  {M{\o}lmer}, \citenamefont {Petrosyan}, \citenamefont {Rabl},\ and\
  \citenamefont {Schmiedmayer}}]{Kurizki2015}%
  \BibitemOpen
  \bibfield  {author} {\bibinfo {author} {\bibfnamefont {G.}~\bibnamefont
  {Kurizki}}, \bibinfo {author} {\bibfnamefont {P.}~\bibnamefont {Bertet}},
  \bibinfo {author} {\bibfnamefont {Y.}~\bibnamefont {Kubo}}, \bibinfo {author}
  {\bibfnamefont {K.}~\bibnamefont {M{\o}lmer}}, \bibinfo {author}
  {\bibfnamefont {D.}~\bibnamefont {Petrosyan}}, \bibinfo {author}
  {\bibfnamefont {P.}~\bibnamefont {Rabl}}, \ and\ \bibinfo {author}
  {\bibfnamefont {J.}~\bibnamefont {Schmiedmayer}},\ }\href {\doibase
  10.1073/pnas.1419326112} {\bibfield  {journal} {\bibinfo  {journal}
  {Proceedings of the National Academy of Sciences}\ }\textbf {\bibinfo
  {volume} {112}},\ \bibinfo {pages} {3866} (\bibinfo {year} {2015})},\ \Eprint
  {http://arxiv.org/abs/https://www.pnas.org/content/112/13/3866.full.pdf}
  {https://www.pnas.org/content/112/13/3866.full.pdf} \BibitemShut {NoStop}%
\bibitem [{\citenamefont {Degen}\ \emph {et~al.}(2017)\citenamefont {Degen},
  \citenamefont {Reinhard},\ and\ \citenamefont {Cappellaro}}]{Degen2017}%
  \BibitemOpen
  \bibfield  {author} {\bibinfo {author} {\bibfnamefont {C.~L.}\ \bibnamefont
  {Degen}}, \bibinfo {author} {\bibfnamefont {F.}~\bibnamefont {Reinhard}}, \
  and\ \bibinfo {author} {\bibfnamefont {P.}~\bibnamefont {Cappellaro}},\
  }\href {\doibase 10.1103/RevModPhys.89.035002} {\bibfield  {journal}
  {\bibinfo  {journal} {Rev. Mod. Phys.}\ }\textbf {\bibinfo {volume} {89}},\
  \bibinfo {pages} {035002} (\bibinfo {year} {2017})}\BibitemShut {NoStop}%
\bibitem [{\citenamefont {{Thiel}}\ \emph {et~al.}(2011)\citenamefont
  {{Thiel}}, \citenamefont {{B{\"o}ttger}},\ and\ \citenamefont
  {{Cone}}}]{Thiel2011Lumin}%
  \BibitemOpen
  \bibfield  {author} {\bibinfo {author} {\bibfnamefont {C.~W.}\ \bibnamefont
  {{Thiel}}}, \bibinfo {author} {\bibfnamefont {T.}~\bibnamefont
  {{B{\"o}ttger}}}, \ and\ \bibinfo {author} {\bibfnamefont {R.~L.}\
  \bibnamefont {{Cone}}},\ }\href {\doibase 10.1016/j.jlumin.2010.12.015}
  {\bibfield  {journal} {\bibinfo  {journal} {Journal of Luminescence}\
  }\textbf {\bibinfo {volume} {131}},\ \bibinfo {pages} {353} (\bibinfo {year}
  {2011})}\BibitemShut {NoStop}%
\bibitem [{\citenamefont {Equall}\ \emph {et~al.}(1994)\citenamefont {Equall},
  \citenamefont {Sun}, \citenamefont {Cone},\ and\ \citenamefont
  {Macfarlane}}]{Equall1994}%
  \BibitemOpen
  \bibfield  {author} {\bibinfo {author} {\bibfnamefont {R.~W.}\ \bibnamefont
  {Equall}}, \bibinfo {author} {\bibfnamefont {Y.}~\bibnamefont {Sun}},
  \bibinfo {author} {\bibfnamefont {R.~L.}\ \bibnamefont {Cone}}, \ and\
  \bibinfo {author} {\bibfnamefont {R.~M.}\ \bibnamefont {Macfarlane}},\ }\href
  {\doibase 10.1103/PhysRevLett.72.2179} {\bibfield  {journal} {\bibinfo
  {journal} {Phys. Rev. Lett.}\ }\textbf {\bibinfo {volume} {72}},\ \bibinfo
  {pages} {2179} (\bibinfo {year} {1994})}\BibitemShut {NoStop}%
\bibitem [{\citenamefont {Zhong}\ \emph {et~al.}(2015)\citenamefont {Zhong},
  \citenamefont {Hedges}, \citenamefont {Ahlefeldt}, \citenamefont
  {Bartholomew}, \citenamefont {Beavan}, \citenamefont {Wittig}, \citenamefont
  {Longdell},\ and\ \citenamefont {Sellars}}]{MZhong2015}%
  \BibitemOpen
  \bibfield  {author} {\bibinfo {author} {\bibfnamefont {M.}~\bibnamefont
  {Zhong}}, \bibinfo {author} {\bibfnamefont {M.~P.}\ \bibnamefont {Hedges}},
  \bibinfo {author} {\bibfnamefont {R.~L.}\ \bibnamefont {Ahlefeldt}}, \bibinfo
  {author} {\bibfnamefont {J.~G.}\ \bibnamefont {Bartholomew}}, \bibinfo
  {author} {\bibfnamefont {S.~E.}\ \bibnamefont {Beavan}}, \bibinfo {author}
  {\bibfnamefont {S.~M.}\ \bibnamefont {Wittig}}, \bibinfo {author}
  {\bibfnamefont {J.~J.}\ \bibnamefont {Longdell}}, \ and\ \bibinfo {author}
  {\bibfnamefont {M.~J.}\ \bibnamefont {Sellars}},\ }\href@noop {} {\bibfield
  {journal} {\bibinfo  {journal} {Nature}\ }\textbf {\bibinfo {volume} {517}},\
  \bibinfo {pages} {177} (\bibinfo {year} {2015})}\BibitemShut {NoStop}%
\bibitem [{\citenamefont {B\"ottger}\ \emph {et~al.}(2009)\citenamefont
  {B\"ottger}, \citenamefont {Thiel}, \citenamefont {Cone},\ and\ \citenamefont
  {Sun}}]{Bottger2009}%
  \BibitemOpen
  \bibfield  {author} {\bibinfo {author} {\bibfnamefont {T.}~\bibnamefont
  {B\"ottger}}, \bibinfo {author} {\bibfnamefont {C.~W.}\ \bibnamefont
  {Thiel}}, \bibinfo {author} {\bibfnamefont {R.~L.}\ \bibnamefont {Cone}}, \
  and\ \bibinfo {author} {\bibfnamefont {Y.}~\bibnamefont {Sun}},\ }\href
  {\doibase 10.1103/PhysRevB.79.115104} {\bibfield  {journal} {\bibinfo
  {journal} {Phys. Rev. B}\ }\textbf {\bibinfo {volume} {79}},\ \bibinfo
  {pages} {115104} (\bibinfo {year} {2009})}\BibitemShut {NoStop}%
\bibitem [{\citenamefont {Rančić}\ \emph {et~al.}(2017)\citenamefont
  {Rančić}, \citenamefont {Hedges}, \citenamefont {Ahlefeldt},\ and\
  \citenamefont {Sellars}}]{Rancic2017}%
  \BibitemOpen
  \bibfield  {author} {\bibinfo {author} {\bibfnamefont {M.}~\bibnamefont
  {Rančić}}, \bibinfo {author} {\bibfnamefont {M.~P.}\ \bibnamefont
  {Hedges}}, \bibinfo {author} {\bibfnamefont {R.~L.}\ \bibnamefont
  {Ahlefeldt}}, \ and\ \bibinfo {author} {\bibfnamefont {M.~J.}\ \bibnamefont
  {Sellars}},\ }\href {\doibase 10.1038/nphys4254} {\bibfield  {journal}
  {\bibinfo  {journal} {Nature Physics}\ }\textbf {\bibinfo {volume} {14}},\
  \bibinfo {pages} {50–54} (\bibinfo {year} {2017})}\BibitemShut {NoStop}%
\bibitem [{\citenamefont {Sun}\ \emph {et~al.}(2008)\citenamefont {Sun},
  \citenamefont {B\"ottger}, \citenamefont {Thiel},\ and\ \citenamefont
  {Cone}}]{Sun2008}%
  \BibitemOpen
  \bibfield  {author} {\bibinfo {author} {\bibfnamefont {Y.}~\bibnamefont
  {Sun}}, \bibinfo {author} {\bibfnamefont {T.}~\bibnamefont {B\"ottger}},
  \bibinfo {author} {\bibfnamefont {C.~W.}\ \bibnamefont {Thiel}}, \ and\
  \bibinfo {author} {\bibfnamefont {R.~L.}\ \bibnamefont {Cone}},\ }\href
  {\doibase 10.1103/PhysRevB.77.085124} {\bibfield  {journal} {\bibinfo
  {journal} {Phys. Rev. B}\ }\textbf {\bibinfo {volume} {77}},\ \bibinfo
  {pages} {085124} (\bibinfo {year} {2008})}\BibitemShut {NoStop}%
\bibitem [{\citenamefont {Harris}(2001)}]{Harris2001}%
  \BibitemOpen
  \bibfield  {author} {\bibinfo {author} {\bibfnamefont {T.~L.}\ \bibnamefont
  {Harris}},\ }\emph {\bibinfo {title} {Erbium-based optical coherent transient
  correlator for the 1.5-micron communication bands}},\ \href@noop {} {Ph.D.
  thesis},\ \bibinfo  {school} {Montana State University}, \bibinfo {address}
  {Bozeman, Montana} (\bibinfo {year} {2001})\BibitemShut {NoStop}%
\bibitem [{\citenamefont {Kukharchyk}\ \emph {et~al.}(2018)\citenamefont
  {Kukharchyk}, \citenamefont {Sholokhov}, \citenamefont {Morozov},
  \citenamefont {Korableva}, \citenamefont {Kalachev},\ and\ \citenamefont
  {Bushev}}]{Kukharchyk2018}%
  \BibitemOpen
  \bibfield  {author} {\bibinfo {author} {\bibfnamefont {N.}~\bibnamefont
  {Kukharchyk}}, \bibinfo {author} {\bibfnamefont {D.}~\bibnamefont
  {Sholokhov}}, \bibinfo {author} {\bibfnamefont {O.}~\bibnamefont {Morozov}},
  \bibinfo {author} {\bibfnamefont {S.}~\bibnamefont {Korableva}}, \bibinfo
  {author} {\bibfnamefont {A.}~\bibnamefont {Kalachev}}, \ and\ \bibinfo
  {author} {\bibfnamefont {P.}~\bibnamefont {Bushev}},\ }\href@noop {}
  {\bibfield  {journal} {\bibinfo  {journal} {New Journal of Physics}\ }\textbf
  {\bibinfo {volume} {20}},\ \bibinfo {pages} {023044} (\bibinfo {year}
  {2018})}\BibitemShut {NoStop}%
\bibitem [{\citenamefont {Craiciu}\ \emph {et~al.}(2019)\citenamefont
  {Craiciu}, \citenamefont {Lei}, \citenamefont {Rochman}, \citenamefont
  {Kindem}, \citenamefont {Bartholomew}, \citenamefont {Miyazono},
  \citenamefont {Zhong}, \citenamefont {Sinclair},\ and\ \citenamefont
  {Faraon}}]{Craiciu2019}%
  \BibitemOpen
  \bibfield  {author} {\bibinfo {author} {\bibfnamefont {I.}~\bibnamefont
  {Craiciu}}, \bibinfo {author} {\bibfnamefont {M.}~\bibnamefont {Lei}},
  \bibinfo {author} {\bibfnamefont {J.}~\bibnamefont {Rochman}}, \bibinfo
  {author} {\bibfnamefont {J.~M.}\ \bibnamefont {Kindem}}, \bibinfo {author}
  {\bibfnamefont {J.~G.}\ \bibnamefont {Bartholomew}}, \bibinfo {author}
  {\bibfnamefont {E.}~\bibnamefont {Miyazono}}, \bibinfo {author}
  {\bibfnamefont {T.}~\bibnamefont {Zhong}}, \bibinfo {author} {\bibfnamefont
  {N.}~\bibnamefont {Sinclair}}, \ and\ \bibinfo {author} {\bibfnamefont
  {A.}~\bibnamefont {Faraon}},\ }\href {\doibase
  10.1103/PhysRevApplied.12.024062} {\bibfield  {journal} {\bibinfo  {journal}
  {Phys. Rev. Applied}\ }\textbf {\bibinfo {volume} {12}},\ \bibinfo {pages}
  {024062} (\bibinfo {year} {2019})}\BibitemShut {NoStop}%
\bibitem [{\citenamefont {B\"ottger}\ \emph {et~al.}(2006)\citenamefont
  {B\"ottger}, \citenamefont {Thiel}, \citenamefont {Sun},\ and\ \citenamefont
  {Cone}}]{Bottger2006}%
  \BibitemOpen
  \bibfield  {author} {\bibinfo {author} {\bibfnamefont {T.}~\bibnamefont
  {B\"ottger}}, \bibinfo {author} {\bibfnamefont {C.~W.}\ \bibnamefont
  {Thiel}}, \bibinfo {author} {\bibfnamefont {Y.}~\bibnamefont {Sun}}, \ and\
  \bibinfo {author} {\bibfnamefont {R.~L.}\ \bibnamefont {Cone}},\ }\href
  {\doibase 10.1103/PhysRevB.73.075101} {\bibfield  {journal} {\bibinfo
  {journal} {Phys. Rev. B}\ }\textbf {\bibinfo {volume} {73}},\ \bibinfo
  {pages} {075101} (\bibinfo {year} {2006})}\BibitemShut {NoStop}%
\bibitem [{\citenamefont {Zhong}\ and\ \citenamefont
  {Goldner}(2019)}]{TZhong2019}%
  \BibitemOpen
  \bibfield  {author} {\bibinfo {author} {\bibfnamefont {T.}~\bibnamefont
  {Zhong}}\ and\ \bibinfo {author} {\bibfnamefont {P.}~\bibnamefont
  {Goldner}},\ }\href {\doibase 10.1515/nanoph-2019-0185} {\bibfield  {journal}
  {\bibinfo  {journal} {Nanophotonics}\ } (\bibinfo {year} {2019}),\
  10.1515/nanoph-2019-0185}\BibitemShut {NoStop}%
\bibitem [{\citenamefont {Scarafagio}\ \emph {et~al.}(2019)\citenamefont
  {Scarafagio}, \citenamefont {Tallaire}, \citenamefont {Tielrooij},
  \citenamefont {Cano}, \citenamefont {Grishin}, \citenamefont {Chavanne},
  \citenamefont {Koppens}, \citenamefont {Ringued{\'e}}, \citenamefont
  {Cassir}, \citenamefont {Serrano}, \citenamefont {Goldner},\ and\
  \citenamefont {Ferrier}}]{Scarafagio2019}%
  \BibitemOpen
  \bibfield  {author} {\bibinfo {author} {\bibfnamefont {M.}~\bibnamefont
  {Scarafagio}}, \bibinfo {author} {\bibfnamefont {A.}~\bibnamefont
  {Tallaire}}, \bibinfo {author} {\bibfnamefont {K.-J.}\ \bibnamefont
  {Tielrooij}}, \bibinfo {author} {\bibfnamefont {D.}~\bibnamefont {Cano}},
  \bibinfo {author} {\bibfnamefont {A.}~\bibnamefont {Grishin}}, \bibinfo
  {author} {\bibfnamefont {M.~H.}\ \bibnamefont {Chavanne}}, \bibinfo {author}
  {\bibfnamefont {F.~H.}\ \bibnamefont {Koppens}}, \bibinfo {author}
  {\bibfnamefont {A.}~\bibnamefont {Ringued{\'e}}}, \bibinfo {author}
  {\bibfnamefont {M.}~\bibnamefont {Cassir}}, \bibinfo {author} {\bibfnamefont
  {D.}~\bibnamefont {Serrano}}, \bibinfo {author} {\bibfnamefont
  {P.}~\bibnamefont {Goldner}}, \ and\ \bibinfo {author} {\bibfnamefont
  {A.}~\bibnamefont {Ferrier}},\ }\href@noop {} {\bibfield  {journal} {\bibinfo
   {journal} {The Journal of Physical Chemistry C}\ } (\bibinfo {year}
  {2019})}\BibitemShut {NoStop}%
\bibitem [{\citenamefont {Bartholomew}\ \emph {et~al.}(2017)\citenamefont
  {Bartholomew}, \citenamefont {de~Oliveira~Lima}, \citenamefont {Ferrier},\
  and\ \citenamefont {Goldner}}]{Bartholomew2017}%
  \BibitemOpen
  \bibfield  {author} {\bibinfo {author} {\bibfnamefont {J.~G.}\ \bibnamefont
  {Bartholomew}}, \bibinfo {author} {\bibfnamefont {K.}~\bibnamefont
  {de~Oliveira~Lima}}, \bibinfo {author} {\bibfnamefont {A.}~\bibnamefont
  {Ferrier}}, \ and\ \bibinfo {author} {\bibfnamefont {P.}~\bibnamefont
  {Goldner}},\ }\href@noop {} {\bibfield  {journal} {\bibinfo  {journal} {Nano
  letters}\ }\textbf {\bibinfo {volume} {17}},\ \bibinfo {pages} {778}
  (\bibinfo {year} {2017})}\BibitemShut {NoStop}%
\bibitem [{\citenamefont {Zhang}\ \emph {et~al.}(2017)\citenamefont {Zhang},
  \citenamefont {Yang}, \citenamefont {Gray}, \citenamefont {Brown},
  \citenamefont {Ketcham}, \citenamefont {Baker}, \citenamefont {Carapella},
  \citenamefont {Davis}, \citenamefont {Arroyo},\ and\ \citenamefont
  {Nolan}}]{Zhang2017}%
  \BibitemOpen
  \bibfield  {author} {\bibinfo {author} {\bibfnamefont {H.}~\bibnamefont
  {Zhang}}, \bibinfo {author} {\bibfnamefont {J.}~\bibnamefont {Yang}},
  \bibinfo {author} {\bibfnamefont {S.}~\bibnamefont {Gray}}, \bibinfo {author}
  {\bibfnamefont {J.~A.}\ \bibnamefont {Brown}}, \bibinfo {author}
  {\bibfnamefont {T.~D.}\ \bibnamefont {Ketcham}}, \bibinfo {author}
  {\bibfnamefont {D.~E.}\ \bibnamefont {Baker}}, \bibinfo {author}
  {\bibfnamefont {A.}~\bibnamefont {Carapella}}, \bibinfo {author}
  {\bibfnamefont {R.~W.}\ \bibnamefont {Davis}}, \bibinfo {author}
  {\bibfnamefont {J.~G.}\ \bibnamefont {Arroyo}}, \ and\ \bibinfo {author}
  {\bibfnamefont {D.~A.}\ \bibnamefont {Nolan}},\ }\href@noop {} {\bibfield
  {journal} {\bibinfo  {journal} {ACS Omega}\ }\textbf {\bibinfo {volume}
  {2}},\ \bibinfo {pages} {3739} (\bibinfo {year} {2017})}\BibitemShut
  {NoStop}%
\bibitem [{\citenamefont {Macfarlane}\ and\ \citenamefont
  {Shelby}(1981)}]{Macfarlane1981}%
  \BibitemOpen
  \bibfield  {author} {\bibinfo {author} {\bibfnamefont {R.}~\bibnamefont
  {Macfarlane}}\ and\ \bibinfo {author} {\bibfnamefont {R.}~\bibnamefont
  {Shelby}},\ }\href {\doibase https://doi.org/10.1016/0030-4018(81)90048-1}
  {\bibfield  {journal} {\bibinfo  {journal} {Optics Communications}\ }\textbf
  {\bibinfo {volume} {39}},\ \bibinfo {pages} {169 } (\bibinfo {year}
  {1981})}\BibitemShut {NoStop}%
\bibitem [{\citenamefont {Perrot}\ \emph {et~al.}(2013)\citenamefont {Perrot},
  \citenamefont {Goldner}, \citenamefont {Giaume}, \citenamefont
  {Lovri\ifmmode~\acute{c}\else \'{c}\fi{}}, \citenamefont {Andriamiadamanana},
  \citenamefont {Gon\ifmmode~\mbox{\c{c}}\else \c{c}\fi{}alves},\ and\
  \citenamefont {Ferrier}}]{Perrot2013}%
  \BibitemOpen
  \bibfield  {author} {\bibinfo {author} {\bibfnamefont {A.}~\bibnamefont
  {Perrot}}, \bibinfo {author} {\bibfnamefont {P.}~\bibnamefont {Goldner}},
  \bibinfo {author} {\bibfnamefont {D.}~\bibnamefont {Giaume}}, \bibinfo
  {author} {\bibfnamefont {M.}~\bibnamefont {Lovri\ifmmode~\acute{c}\else
  \'{c}\fi{}}}, \bibinfo {author} {\bibfnamefont {C.}~\bibnamefont
  {Andriamiadamanana}}, \bibinfo {author} {\bibfnamefont {R.~R.}\ \bibnamefont
  {Gon\ifmmode~\mbox{\c{c}}\else \c{c}\fi{}alves}}, \ and\ \bibinfo {author}
  {\bibfnamefont {A.}~\bibnamefont {Ferrier}},\ }\href {\doibase
  10.1103/PhysRevLett.111.203601} {\bibfield  {journal} {\bibinfo  {journal}
  {Phys. Rev. Lett.}\ }\textbf {\bibinfo {volume} {111}},\ \bibinfo {pages}
  {203601} (\bibinfo {year} {2013})}\BibitemShut {NoStop}%
\bibitem [{\citenamefont {Kunkel}\ \emph {et~al.}(2016)\citenamefont {Kunkel},
  \citenamefont {Bartholomew}, \citenamefont {Welinski}, \citenamefont
  {Ferrier}, \citenamefont {Ikesue},\ and\ \citenamefont
  {Goldner}}]{Kunkel2016}%
  \BibitemOpen
  \bibfield  {author} {\bibinfo {author} {\bibfnamefont {N.}~\bibnamefont
  {Kunkel}}, \bibinfo {author} {\bibfnamefont {J.}~\bibnamefont {Bartholomew}},
  \bibinfo {author} {\bibfnamefont {S.}~\bibnamefont {Welinski}}, \bibinfo
  {author} {\bibfnamefont {A.}~\bibnamefont {Ferrier}}, \bibinfo {author}
  {\bibfnamefont {A.}~\bibnamefont {Ikesue}}, \ and\ \bibinfo {author}
  {\bibfnamefont {P.}~\bibnamefont {Goldner}},\ }\href {\doibase
  10.1103/PhysRevB.94.184301} {\bibfield  {journal} {\bibinfo  {journal} {Phys.
  Rev. B}\ }\textbf {\bibinfo {volume} {94}},\ \bibinfo {pages} {184301}
  (\bibinfo {year} {2016})}\BibitemShut {NoStop}%
\bibitem [{\citenamefont {Ferrier}\ \emph {et~al.}(2013)\citenamefont
  {Ferrier}, \citenamefont {Thiel}, \citenamefont {Tumino}, \citenamefont
  {Ramirez}, \citenamefont {Baus\'a}, \citenamefont {Cone}, \citenamefont
  {Ikesue},\ and\ \citenamefont {Goldner}}]{Ferrier2013}%
  \BibitemOpen
  \bibfield  {author} {\bibinfo {author} {\bibfnamefont {A.}~\bibnamefont
  {Ferrier}}, \bibinfo {author} {\bibfnamefont {C.~W.}\ \bibnamefont {Thiel}},
  \bibinfo {author} {\bibfnamefont {B.}~\bibnamefont {Tumino}}, \bibinfo
  {author} {\bibfnamefont {M.~O.}\ \bibnamefont {Ramirez}}, \bibinfo {author}
  {\bibfnamefont {L.~E.}\ \bibnamefont {Baus\'a}}, \bibinfo {author}
  {\bibfnamefont {R.~L.}\ \bibnamefont {Cone}}, \bibinfo {author}
  {\bibfnamefont {A.}~\bibnamefont {Ikesue}}, \ and\ \bibinfo {author}
  {\bibfnamefont {P.}~\bibnamefont {Goldner}},\ }\href {\doibase
  10.1103/PhysRevB.87.041102} {\bibfield  {journal} {\bibinfo  {journal} {Phys.
  Rev. B}\ }\textbf {\bibinfo {volume} {87}},\ \bibinfo {pages} {041102(R)}
  (\bibinfo {year} {2013})}\BibitemShut {NoStop}%
\bibitem [{\citenamefont {Bartolo}(1968)}]{Bartolo1968}%
  \BibitemOpen
  \bibfield  {author} {\bibinfo {author} {\bibfnamefont {B.~D.}\ \bibnamefont
  {Bartolo}},\ }\href@noop {} {\emph {\bibinfo {title} {Optical interactions in
  solids}}}\ (\bibinfo  {publisher} {Wiley},\ \bibinfo {year}
  {1968})\BibitemShut {NoStop}%
\bibitem [{\citenamefont {Henderson}\ and\ \citenamefont
  {Imbusch}(2006)}]{Henderson2006}%
  \BibitemOpen
  \bibfield  {author} {\bibinfo {author} {\bibfnamefont {B.}~\bibnamefont
  {Henderson}}\ and\ \bibinfo {author} {\bibfnamefont {G.~F.}\ \bibnamefont
  {Imbusch}},\ }\href@noop {} {\emph {\bibinfo {title} {Optical Spectroscopy of
  Inorganic Solids}}}\ (\bibinfo  {publisher} {Oxford University Press},\
  \bibinfo {year} {2006})\BibitemShut {NoStop}%
\bibitem [{\citenamefont {Liu}\ and\ \citenamefont {Jacquier}(2010)}]{Liu2010}%
  \BibitemOpen
  \bibinfo {editor} {\bibfnamefont {G.}~\bibnamefont {Liu}}\ and\ \bibinfo
  {editor} {\bibfnamefont {B.}~\bibnamefont {Jacquier}},\ eds.,\ \href@noop {}
  {\emph {\bibinfo {title} {Spectroscopic properties of rare earths in optical
  materials}}}\ (\bibinfo  {publisher} {Tsinghua Univ. Press},\ \bibinfo {year}
  {2010})\BibitemShut {NoStop}%
\bibitem [{\citenamefont {Mims}(1968)}]{Mims1968}%
  \BibitemOpen
  \bibfield  {author} {\bibinfo {author} {\bibfnamefont {W.~B.}\ \bibnamefont
  {Mims}},\ }\href {\doibase 10.1103/PhysRev.168.370} {\bibfield  {journal}
  {\bibinfo  {journal} {Phys. Rev.}\ }\textbf {\bibinfo {volume} {168}},\
  \bibinfo {pages} {370} (\bibinfo {year} {1968})}\BibitemShut {NoStop}%
\bibitem [{\citenamefont {Liu}\ and\ \citenamefont {Cone}(1990)}]{Liu1990}%
  \BibitemOpen
  \bibfield  {author} {\bibinfo {author} {\bibfnamefont {G.~K.}\ \bibnamefont
  {Liu}}\ and\ \bibinfo {author} {\bibfnamefont {R.~L.}\ \bibnamefont {Cone}},\
  }\href {\doibase 10.1103/PhysRevB.41.6193} {\bibfield  {journal} {\bibinfo
  {journal} {Phys. Rev. B}\ }\textbf {\bibinfo {volume} {41}},\ \bibinfo
  {pages} {6193} (\bibinfo {year} {1990})}\BibitemShut {NoStop}%
\bibitem [{\citenamefont {Car}\ \emph {et~al.}(2018)\citenamefont {Car},
  \citenamefont {Veissier}, \citenamefont {Louchet-Chauvet}, \citenamefont
  {Le~Gou\"et},\ and\ \citenamefont {Chaneli\`ere}}]{Car2018}%
  \BibitemOpen
  \bibfield  {author} {\bibinfo {author} {\bibfnamefont {B.}~\bibnamefont
  {Car}}, \bibinfo {author} {\bibfnamefont {L.}~\bibnamefont {Veissier}},
  \bibinfo {author} {\bibfnamefont {A.}~\bibnamefont {Louchet-Chauvet}},
  \bibinfo {author} {\bibfnamefont {J.-L.}\ \bibnamefont {Le~Gou\"et}}, \ and\
  \bibinfo {author} {\bibfnamefont {T.}~\bibnamefont {Chaneli\`ere}},\ }\href
  {\doibase 10.1103/PhysRevLett.120.197401} {\bibfield  {journal} {\bibinfo
  {journal} {Phys. Rev. Lett.}\ }\textbf {\bibinfo {volume} {120}},\ \bibinfo
  {pages} {197401} (\bibinfo {year} {2018})}\BibitemShut {NoStop}%
\bibitem [{\citenamefont {Anderson}\ \emph {et~al.}(1972)\citenamefont
  {Anderson}, \citenamefont {Halperin},\ and\ \citenamefont
  {Varma}}]{Anderson1972}%
  \BibitemOpen
  \bibfield  {author} {\bibinfo {author} {\bibfnamefont {P.~W.}\ \bibnamefont
  {Anderson}}, \bibinfo {author} {\bibfnamefont {B.~I.}\ \bibnamefont
  {Halperin}}, \ and\ \bibinfo {author} {\bibfnamefont {C.~M.}\ \bibnamefont
  {Varma}},\ }\href {\doibase 10.1080/14786437208229210} {\bibfield  {journal}
  {\bibinfo  {journal} {The Philosophical Magazine: A Journal of Theoretical
  Experimental and Applied Physics}\ }\textbf {\bibinfo {volume} {25}},\
  \bibinfo {pages} {1} (\bibinfo {year} {1972})}\BibitemShut {NoStop}%
\bibitem [{\citenamefont {Phillips}(1972)}]{Phillips1972}%
  \BibitemOpen
  \bibfield  {author} {\bibinfo {author} {\bibfnamefont {W.~A.}\ \bibnamefont
  {Phillips}},\ }\href {\doibase 10.1007/BF00660072} {\bibfield  {journal}
  {\bibinfo  {journal} {Journal of Low Temperature Physics}\ }\textbf {\bibinfo
  {volume} {7}},\ \bibinfo {pages} {351} (\bibinfo {year} {1972})}\BibitemShut
  {NoStop}%
\bibitem [{\citenamefont {Macfarlane}\ \emph {et~al.}(2007)\citenamefont
  {Macfarlane}, \citenamefont {Sun}, \citenamefont {Sellin},\ and\
  \citenamefont {Cone}}]{Macfarlane2007}%
  \BibitemOpen
  \bibfield  {author} {\bibinfo {author} {\bibfnamefont {R.}~\bibnamefont
  {Macfarlane}}, \bibinfo {author} {\bibfnamefont {Y.}~\bibnamefont {Sun}},
  \bibinfo {author} {\bibfnamefont {P.}~\bibnamefont {Sellin}}, \ and\ \bibinfo
  {author} {\bibfnamefont {R.}~\bibnamefont {Cone}},\ }\href {\doibase
  https://doi.org/10.1016/j.jlumin.2007.02.060} {\bibfield  {journal} {\bibinfo
   {journal} {Journal of Luminescence}\ }\textbf {\bibinfo {volume} {127}},\
  \bibinfo {pages} {61 } (\bibinfo {year} {2007})}\BibitemShut {NoStop}%
\bibitem [{\citenamefont {Huber}\ \emph {et~al.}(1984)\citenamefont {Huber},
  \citenamefont {Broer},\ and\ \citenamefont {Golding}}]{Huber1984}%
  \BibitemOpen
  \bibfield  {author} {\bibinfo {author} {\bibfnamefont {D.~L.}\ \bibnamefont
  {Huber}}, \bibinfo {author} {\bibfnamefont {M.~M.}\ \bibnamefont {Broer}}, \
  and\ \bibinfo {author} {\bibfnamefont {B.}~\bibnamefont {Golding}},\ }\href
  {\doibase 10.1103/PhysRevLett.52.2281} {\bibfield  {journal} {\bibinfo
  {journal} {Phys. Rev. Lett.}\ }\textbf {\bibinfo {volume} {52}},\ \bibinfo
  {pages} {2281} (\bibinfo {year} {1984})}\BibitemShut {NoStop}%
\bibitem [{\citenamefont {Black}\ and\ \citenamefont
  {Halperin}(1977)}]{Black1977}%
  \BibitemOpen
  \bibfield  {author} {\bibinfo {author} {\bibfnamefont {J.~L.}\ \bibnamefont
  {Black}}\ and\ \bibinfo {author} {\bibfnamefont {B.~I.}\ \bibnamefont
  {Halperin}},\ }\href {\doibase 10.1103/PhysRevB.16.2879} {\bibfield
  {journal} {\bibinfo  {journal} {Phys. Rev. B}\ }\textbf {\bibinfo {volume}
  {16}},\ \bibinfo {pages} {2879} (\bibinfo {year} {1977})}\BibitemShut
  {NoStop}%
\bibitem [{\citenamefont {Breinl}\ \emph {et~al.}(1984)\citenamefont {Breinl},
  \citenamefont {Friedrich},\ and\ \citenamefont {Haarer}}]{Breinl1984}%
  \BibitemOpen
  \bibfield  {author} {\bibinfo {author} {\bibfnamefont {W.}~\bibnamefont
  {Breinl}}, \bibinfo {author} {\bibfnamefont {J.}~\bibnamefont {Friedrich}}, \
  and\ \bibinfo {author} {\bibfnamefont {D.}~\bibnamefont {Haarer}},\ }\href
  {\doibase 10.1063/1.448184} {\bibfield  {journal} {\bibinfo  {journal} {The
  Journal of Chemical Physics}\ }\textbf {\bibinfo {volume} {81}},\ \bibinfo
  {pages} {3915} (\bibinfo {year} {1984})}\BibitemShut {NoStop}%
\bibitem [{\citenamefont {Sun}\ \emph {et~al.}(2012)\citenamefont {Sun},
  \citenamefont {Thiel},\ and\ \citenamefont {Cone}}]{Sun2012}%
  \BibitemOpen
  \bibfield  {author} {\bibinfo {author} {\bibfnamefont {Y.}~\bibnamefont
  {Sun}}, \bibinfo {author} {\bibfnamefont {C.~W.}\ \bibnamefont {Thiel}}, \
  and\ \bibinfo {author} {\bibfnamefont {R.~L.}\ \bibnamefont {Cone}},\ }\href
  {\doibase 10.1103/PhysRevB.85.165106} {\bibfield  {journal} {\bibinfo
  {journal} {Phys. Rev. B}\ }\textbf {\bibinfo {volume} {85}},\ \bibinfo
  {pages} {165106} (\bibinfo {year} {2012})}\BibitemShut {NoStop}%
\bibitem [{\citenamefont {Flinn}\ \emph {et~al.}(1994)\citenamefont {Flinn},
  \citenamefont {Jang}, \citenamefont {Ganem}, \citenamefont {Jones},
  \citenamefont {Meltzer},\ and\ \citenamefont {Macfarlane}}]{Flinn1994}%
  \BibitemOpen
  \bibfield  {author} {\bibinfo {author} {\bibfnamefont {G.~P.}\ \bibnamefont
  {Flinn}}, \bibinfo {author} {\bibfnamefont {K.~W.}\ \bibnamefont {Jang}},
  \bibinfo {author} {\bibfnamefont {J.}~\bibnamefont {Ganem}}, \bibinfo
  {author} {\bibfnamefont {M.~L.}\ \bibnamefont {Jones}}, \bibinfo {author}
  {\bibfnamefont {R.~S.}\ \bibnamefont {Meltzer}}, \ and\ \bibinfo {author}
  {\bibfnamefont {R.~M.}\ \bibnamefont {Macfarlane}},\ }\href {\doibase
  10.1103/PhysRevB.49.5821} {\bibfield  {journal} {\bibinfo  {journal} {Phys.
  Rev. B}\ }\textbf {\bibinfo {volume} {49}},\ \bibinfo {pages} {5821}
  (\bibinfo {year} {1994})}\BibitemShut {NoStop}%
\bibitem [{\citenamefont {Macfarlane}\ \emph {et~al.}(2004)\citenamefont
  {Macfarlane}, \citenamefont {Sun}, \citenamefont {Cone}, \citenamefont
  {Thiel},\ and\ \citenamefont {Equall}}]{Macfarlane2004}%
  \BibitemOpen
  \bibfield  {author} {\bibinfo {author} {\bibfnamefont {R.}~\bibnamefont
  {Macfarlane}}, \bibinfo {author} {\bibfnamefont {Y.}~\bibnamefont {Sun}},
  \bibinfo {author} {\bibfnamefont {R.}~\bibnamefont {Cone}}, \bibinfo {author}
  {\bibfnamefont {C.}~\bibnamefont {Thiel}}, \ and\ \bibinfo {author}
  {\bibfnamefont {R.}~\bibnamefont {Equall}},\ }\href {\doibase
  https://doi.org/10.1016/j.jlumin.2003.12.029} {\bibfield  {journal} {\bibinfo
   {journal} {Journal of Luminescence}\ }\textbf {\bibinfo {volume} {107}},\
  \bibinfo {pages} {310 } (\bibinfo {year} {2004})},\ \bibinfo {note}
  {proceedings of the 8th International Meeting on Hole Burning, Single
  Molecule, and Related Spectroscopies: Science and Applications}\BibitemShut
  {NoStop}%
\bibitem [{\citenamefont {Ding}\ \emph {et~al.}(2018)\citenamefont {Ding},
  \citenamefont {van Driel}, \citenamefont {Pereira}, \citenamefont {Bauters},
  \citenamefont {Heck}, \citenamefont {Welker}, \citenamefont {de~Dood},
  \citenamefont {Vantomme}, \citenamefont {Bowers}, \citenamefont {Löffler},\
  and\ \citenamefont {Bouwmeester}}]{Ding2018}%
  \BibitemOpen
  \bibfield  {author} {\bibinfo {author} {\bibfnamefont {D.}~\bibnamefont
  {Ding}}, \bibinfo {author} {\bibfnamefont {D.}~\bibnamefont {van Driel}},
  \bibinfo {author} {\bibfnamefont {L.~M.~C.}\ \bibnamefont {Pereira}},
  \bibinfo {author} {\bibfnamefont {J.~F.}\ \bibnamefont {Bauters}}, \bibinfo
  {author} {\bibfnamefont {M.~J.~R.}\ \bibnamefont {Heck}}, \bibinfo {author}
  {\bibfnamefont {G.}~\bibnamefont {Welker}}, \bibinfo {author} {\bibfnamefont
  {M.~J.~A.}\ \bibnamefont {de~Dood}}, \bibinfo {author} {\bibfnamefont
  {A.}~\bibnamefont {Vantomme}}, \bibinfo {author} {\bibfnamefont {J.~E.}\
  \bibnamefont {Bowers}}, \bibinfo {author} {\bibfnamefont {W.}~\bibnamefont
  {Löffler}}, \ and\ \bibinfo {author} {\bibfnamefont {D.}~\bibnamefont
  {Bouwmeester}},\ }\href@noop {} {\enquote {\bibinfo {title} {Probing
  interacting two-level systems with rare-earth ions},}\ } (\bibinfo {year}
  {2018}),\ \Eprint {http://arxiv.org/abs/1811.05248} {arXiv:1811.05248
  [cond-mat.dis-nn]} \BibitemShut {NoStop}%
\bibitem [{\citenamefont {Staudt}\ \emph {et~al.}(2006)\citenamefont {Staudt},
  \citenamefont {Hastings-Simon}, \citenamefont {Afzelius}, \citenamefont
  {Jaccard}, \citenamefont {Tittel},\ and\ \citenamefont {Gisin}}]{Staudt2006}%
  \BibitemOpen
  \bibfield  {author} {\bibinfo {author} {\bibfnamefont {M.~U.}\ \bibnamefont
  {Staudt}}, \bibinfo {author} {\bibfnamefont {S.~R.}\ \bibnamefont
  {Hastings-Simon}}, \bibinfo {author} {\bibfnamefont {M.}~\bibnamefont
  {Afzelius}}, \bibinfo {author} {\bibfnamefont {D.}~\bibnamefont {Jaccard}},
  \bibinfo {author} {\bibfnamefont {W.}~\bibnamefont {Tittel}}, \ and\ \bibinfo
  {author} {\bibfnamefont {N.}~\bibnamefont {Gisin}},\ }\href {\doibase
  https://doi.org/10.1016/j.optcom.2006.05.007} {\bibfield  {journal} {\bibinfo
   {journal} {Optics Communications}\ }\textbf {\bibinfo {volume} {266}},\
  \bibinfo {pages} {720} (\bibinfo {year} {2006})}\BibitemShut {NoStop}%
\bibitem [{\citenamefont {Littau}\ \emph {et~al.}(1992)\citenamefont {Littau},
  \citenamefont {Dugan}, \citenamefont {Chen},\ and\ \citenamefont
  {Fayer}}]{Littau1992}%
  \BibitemOpen
  \bibfield  {author} {\bibinfo {author} {\bibfnamefont {K.~A.}\ \bibnamefont
  {Littau}}, \bibinfo {author} {\bibfnamefont {M.~A.}\ \bibnamefont {Dugan}},
  \bibinfo {author} {\bibfnamefont {S.}~\bibnamefont {Chen}}, \ and\ \bibinfo
  {author} {\bibfnamefont {M.~D.}\ \bibnamefont {Fayer}},\ }\href {\doibase
  10.1063/1.461902} {\bibfield  {journal} {\bibinfo  {journal} {The Journal of
  Chemical Physics}\ }\textbf {\bibinfo {volume} {96}},\ \bibinfo {pages}
  {3484} (\bibinfo {year} {1992})}\BibitemShut {NoStop}%
\bibitem [{\citenamefont {Silbey}\ \emph {et~al.}(1996)\citenamefont {Silbey},
  \citenamefont {Koedijk},\ and\ \citenamefont {V{\"o}lker}}]{Silbey1996}%
  \BibitemOpen
  \bibfield  {author} {\bibinfo {author} {\bibfnamefont {R.}~\bibnamefont
  {Silbey}}, \bibinfo {author} {\bibfnamefont {J.}~\bibnamefont {Koedijk}}, \
  and\ \bibinfo {author} {\bibfnamefont {S.}~\bibnamefont {V{\"o}lker}},\
  }\href@noop {} {\bibfield  {journal} {\bibinfo  {journal} {The Journal of
  chemical physics}\ }\textbf {\bibinfo {volume} {105}},\ \bibinfo {pages}
  {901} (\bibinfo {year} {1996})}\BibitemShut {NoStop}%
\bibitem [{\citenamefont {Koedijk}\ \emph {et~al.}(1996)\citenamefont
  {Koedijk}, \citenamefont {Wannemacher}, \citenamefont {Silbey},\ and\
  \citenamefont {V{\"o}lker}}]{Koedijk1996}%
  \BibitemOpen
  \bibfield  {author} {\bibinfo {author} {\bibfnamefont {J.}~\bibnamefont
  {Koedijk}}, \bibinfo {author} {\bibfnamefont {R.}~\bibnamefont
  {Wannemacher}}, \bibinfo {author} {\bibfnamefont {R.}~\bibnamefont {Silbey}},
  \ and\ \bibinfo {author} {\bibfnamefont {S.}~\bibnamefont {V{\"o}lker}},\
  }\href@noop {} {\bibfield  {journal} {\bibinfo  {journal} {The Journal of
  Physical Chemistry}\ }\textbf {\bibinfo {volume} {100}},\ \bibinfo {pages}
  {19945} (\bibinfo {year} {1996})}\BibitemShut {NoStop}%
\bibitem [{\citenamefont {Veissier}\ \emph {et~al.}(2016)\citenamefont
  {Veissier}, \citenamefont {Falamarzi}, \citenamefont {Lutz}, \citenamefont
  {Saglamyurek}, \citenamefont {Thiel}, \citenamefont {Cone},\ and\
  \citenamefont {Tittel}}]{Veissier2016}%
  \BibitemOpen
  \bibfield  {author} {\bibinfo {author} {\bibfnamefont {L.}~\bibnamefont
  {Veissier}}, \bibinfo {author} {\bibfnamefont {M.}~\bibnamefont {Falamarzi}},
  \bibinfo {author} {\bibfnamefont {T.}~\bibnamefont {Lutz}}, \bibinfo {author}
  {\bibfnamefont {E.}~\bibnamefont {Saglamyurek}}, \bibinfo {author}
  {\bibfnamefont {C.~W.}\ \bibnamefont {Thiel}}, \bibinfo {author}
  {\bibfnamefont {R.~L.}\ \bibnamefont {Cone}}, \ and\ \bibinfo {author}
  {\bibfnamefont {W.}~\bibnamefont {Tittel}},\ }\href {\doibase
  10.1103/PhysRevB.94.195138} {\bibfield  {journal} {\bibinfo  {journal} {Phys.
  Rev. B}\ }\textbf {\bibinfo {volume} {94}},\ \bibinfo {pages} {195138}
  (\bibinfo {year} {2016})}\BibitemShut {NoStop}%
\bibitem [{\citenamefont {Lutz}\ \emph {et~al.}(2017)\citenamefont {Lutz},
  \citenamefont {Veissier}, \citenamefont {Thiel}, \citenamefont {Woodburn},
  \citenamefont {Cone}, \citenamefont {Barclay},\ and\ \citenamefont
  {Tittel}}]{Lutz2017}%
  \BibitemOpen
  \bibfield  {author} {\bibinfo {author} {\bibfnamefont {T.}~\bibnamefont
  {Lutz}}, \bibinfo {author} {\bibfnamefont {L.}~\bibnamefont {Veissier}},
  \bibinfo {author} {\bibfnamefont {C.~W.}\ \bibnamefont {Thiel}}, \bibinfo
  {author} {\bibfnamefont {P.~J.}\ \bibnamefont {Woodburn}}, \bibinfo {author}
  {\bibfnamefont {R.~L.}\ \bibnamefont {Cone}}, \bibinfo {author}
  {\bibfnamefont {P.~E.}\ \bibnamefont {Barclay}}, \ and\ \bibinfo {author}
  {\bibfnamefont {W.}~\bibnamefont {Tittel}},\ }\href@noop {} {\bibfield
  {journal} {\bibinfo  {journal} {Journal of Luminescence}\ }\textbf {\bibinfo
  {volume} {191}},\ \bibinfo {pages} {2} (\bibinfo {year} {2017})}\BibitemShut
  {NoStop}%
\bibitem [{\citenamefont {Lutz}\ \emph {et~al.}(2016)\citenamefont {Lutz},
  \citenamefont {Veissier}, \citenamefont {Thiel}, \citenamefont {Cone},
  \citenamefont {Barclay},\ and\ \citenamefont {Tittel}}]{Lutz2016}%
  \BibitemOpen
  \bibfield  {author} {\bibinfo {author} {\bibfnamefont {T.}~\bibnamefont
  {Lutz}}, \bibinfo {author} {\bibfnamefont {L.}~\bibnamefont {Veissier}},
  \bibinfo {author} {\bibfnamefont {C.~W.}\ \bibnamefont {Thiel}}, \bibinfo
  {author} {\bibfnamefont {R.~L.}\ \bibnamefont {Cone}}, \bibinfo {author}
  {\bibfnamefont {P.~E.}\ \bibnamefont {Barclay}}, \ and\ \bibinfo {author}
  {\bibfnamefont {W.}~\bibnamefont {Tittel}},\ }\href {\doibase
  10.1103/PhysRevA.94.013801} {\bibfield  {journal} {\bibinfo  {journal} {Phys.
  Rev. A}\ }\textbf {\bibinfo {volume} {94}},\ \bibinfo {pages} {013801}
  (\bibinfo {year} {2016})}\BibitemShut {NoStop}%
\bibitem [{\citenamefont {Afzelius}\ \emph {et~al.}(2009)\citenamefont
  {Afzelius}, \citenamefont {Simon}, \citenamefont {de~Riedmatten},\ and\
  \citenamefont {Gisin}}]{Afzelius2009}%
  \BibitemOpen
  \bibfield  {author} {\bibinfo {author} {\bibfnamefont {M.}~\bibnamefont
  {Afzelius}}, \bibinfo {author} {\bibfnamefont {C.}~\bibnamefont {Simon}},
  \bibinfo {author} {\bibfnamefont {H.}~\bibnamefont {de~Riedmatten}}, \ and\
  \bibinfo {author} {\bibfnamefont {N.}~\bibnamefont {Gisin}},\ }\href
  {\doibase 10.1103/PhysRevA.79.052329} {\bibfield  {journal} {\bibinfo
  {journal} {Phys. Rev. A}\ }\textbf {\bibinfo {volume} {79}},\ \bibinfo
  {pages} {052329} (\bibinfo {year} {2009})}\BibitemShut {NoStop}%
\bibitem [{\citenamefont {Hedges}\ \emph {et~al.}(2010)\citenamefont {Hedges},
  \citenamefont {Longdell}, \citenamefont {Li},\ and\ \citenamefont
  {Sellars}}]{Hedges2010}%
  \BibitemOpen
  \bibfield  {author} {\bibinfo {author} {\bibfnamefont {M.~P.}\ \bibnamefont
  {Hedges}}, \bibinfo {author} {\bibfnamefont {J.~J.}\ \bibnamefont
  {Longdell}}, \bibinfo {author} {\bibfnamefont {Y.}~\bibnamefont {Li}}, \ and\
  \bibinfo {author} {\bibfnamefont {M.~J.}\ \bibnamefont {Sellars}},\
  }\href@noop {} {\bibfield  {journal} {\bibinfo  {journal} {Nature}\ }\textbf
  {\bibinfo {volume} {465}},\ \bibinfo {pages} {1052} (\bibinfo {year}
  {2010})}\BibitemShut {NoStop}%
\bibitem [{\citenamefont {Zhong}\ \emph {et~al.}(2018)\citenamefont {Zhong},
  \citenamefont {Kindem}, \citenamefont {Bartholomew}, \citenamefont {Rochman},
  \citenamefont {Craiciu}, \citenamefont {Verma}, \citenamefont {Nam},
  \citenamefont {Marsili}, \citenamefont {Shaw}, \citenamefont {Beyer},\ and\
  \citenamefont {Faraon}}]{Zhong2018}%
  \BibitemOpen
  \bibfield  {author} {\bibinfo {author} {\bibfnamefont {T.}~\bibnamefont
  {Zhong}}, \bibinfo {author} {\bibfnamefont {J.~M.}\ \bibnamefont {Kindem}},
  \bibinfo {author} {\bibfnamefont {J.~G.}\ \bibnamefont {Bartholomew}},
  \bibinfo {author} {\bibfnamefont {J.}~\bibnamefont {Rochman}}, \bibinfo
  {author} {\bibfnamefont {I.}~\bibnamefont {Craiciu}}, \bibinfo {author}
  {\bibfnamefont {V.}~\bibnamefont {Verma}}, \bibinfo {author} {\bibfnamefont
  {S.~W.}\ \bibnamefont {Nam}}, \bibinfo {author} {\bibfnamefont
  {F.}~\bibnamefont {Marsili}}, \bibinfo {author} {\bibfnamefont {M.~D.}\
  \bibnamefont {Shaw}}, \bibinfo {author} {\bibfnamefont {A.~D.}\ \bibnamefont
  {Beyer}}, \ and\ \bibinfo {author} {\bibfnamefont {A.}~\bibnamefont
  {Faraon}},\ }\href {\doibase 10.1103/PhysRevLett.121.183603} {\bibfield
  {journal} {\bibinfo  {journal} {Phys. Rev. Lett.}\ }\textbf {\bibinfo
  {volume} {121}},\ \bibinfo {pages} {183603} (\bibinfo {year}
  {2018})}\BibitemShut {NoStop}%
\bibitem [{\citenamefont {Dibos}\ \emph {et~al.}(2018)\citenamefont {Dibos},
  \citenamefont {Raha}, \citenamefont {Phenicie},\ and\ \citenamefont
  {Thompson}}]{Dibos2018}%
  \BibitemOpen
  \bibfield  {author} {\bibinfo {author} {\bibfnamefont {A.~M.}\ \bibnamefont
  {Dibos}}, \bibinfo {author} {\bibfnamefont {M.}~\bibnamefont {Raha}},
  \bibinfo {author} {\bibfnamefont {C.~M.}\ \bibnamefont {Phenicie}}, \ and\
  \bibinfo {author} {\bibfnamefont {J.~D.}\ \bibnamefont {Thompson}},\ }\href
  {\doibase 10.1103/PhysRevLett.120.243601} {\bibfield  {journal} {\bibinfo
  {journal} {Phys. Rev. Lett.}\ }\textbf {\bibinfo {volume} {120}},\ \bibinfo
  {pages} {243601} (\bibinfo {year} {2018})}\BibitemShut {NoStop}%
\end{thebibliography}%

\end{document}